\documentclass[journal,final,twocolumn]{IEEEtran}

\usepackage{graphicx}
\usepackage{subcaption}
\usepackage[cmex10]{amsmath}
\usepackage[misc]{ifsym} % for quincunx dice face symbol
\usepackage{amssymb}
\usepackage{amsthm}
\usepackage[noend]{algpseudocode} % algorithm environment
\usepackage{algorithm}  % float wrapper for algorithms
\usepackage{cite}
\usepackage{xifthen}
\usepackage{flushend}  % for evening out columns on last page
\usepackage{verbatim} % for comment environment

\usepackage{mathdefs}

\newcommand{\sir}{\mathsf{sir}}
\newcommand{\siro}[1]{\sir\!\left(#1\right)}
\newcommand{\sinr}{\mathsf{sinr}}
\newcommand{\sinro}[1]{\sinr\!\left(#1\right)}

\newtheorem{corollary}{Corollary}
\newtheorem{definition}{Definition}

\newtheorem{lemma}{Lemma}
\newtheorem{proposition}{Proposition}

\newtheorem{theorem}{Theorem}

\newcommand{\corref}[1]{Cor.~\ref{cor:#1}}
\newcommand{\defref}[1]{Def.~\ref{def:#1}}

\newcommand{\lemref}[1]{Lem.~\ref{lem:#1}}
\newcommand{\prpref}[1]{Prop.~\ref{prp:#1}}
\newcommand{\thmref}[1]{Thm.~\ref{thm:#1}}
\renewcommand{\eqref}[1]{(\ref{eq:#1})}
\newcommand{\secref}[1]{\S\ref{sec:#1}}

\newcommand{\figref}[1]{Fig.~\ref{fig:#1}}
\newcommand{\tabref}[1]{Table~\ref{tab:#1}}

\begin{document}

\title{On the joint impact of bias and power control on downlink spectral efficiency in cellular networks}

\author{%
Lex~Fridman,~\IEEEmembership{Member,~IEEE},~Jeffrey~Wildman,~\IEEEmembership{Member,~IEEE},~Steven~Weber,~\IEEEmembership{Senior~Member,~IEEE}%
\thanks{Support from the National Science Foundation (awards CNS-1147838 and CNS-1457306) is gratefully acknowledged. A preliminary version of this work was presented at CROWNCOM 2013 \cite{FriWil2013}.  S.\ Weber is the contact author  (\textsf{sweber@coe.drexel.edu}).}%
}

\maketitle

\begin{abstract}
Cell biasing and downlink transmit power are two controls that may be used to improve the spectral efficiency of cellular networks.  With cell biasing, each mobile user associates with the base station offering, say, the highest biased signal to interference plus noise ratio.  Biasing affects the cell association decisions of mobile users, but not the received instantaneous downlink transmission rates.  Adjusting the collection of downlink transmission powers can likewise affect the cell associations, but in contrast with  biasing, it also directly affects the instantaneous rates.  This paper investigates the joint use of both cell biasing and transmission power control and their (individual and joint) effects on the statistical properties of the collection of per-user spectral efficiencies.  Our analytical results and numerical investigations demonstrate in some cases a significant performance improvement in the Pareto efficient frontiers of both a mean-variance and throughput-fairness tradeoff from using both bias and power controls over using either control alone.  
\end{abstract}


%=========================================================================================
\section{Introduction}
\label{sec:intro}

It is envisioned that heterogeneous cellular networks \cite{And2013, GhoManRat2012}, an integration of multiple cellular access technologies, each suited for various data rates, mobility, coverage areas, will enable cellular systems to support both increased user density and service rates through improved spectral efficiency (SE).  Cell biasing and downlink transmit power are two controls that may be used to improve the (downlink) SE of cellular networks.  With cell biasing, each mobile user (MU) associates with the base station (BS) offering, say, the highest biased signal to interference plus noise ratio (SINR).  Biasing affects the cell associations, but not the received instantaneous downlink transmission rates.  Adjusting the transmission powers can likewise affect the cell associations, but in contrast with biasing, it also directly affects the instantaneous rates.   

We suppose the locations of the BSs are fixed, the cell bias parameters and dowlink transmission powers are controls, and the locations of the $m$ MUs are independent and placed uniformly at random in the arena.  This paper investigates the joint use of both cell biasing and transmission power controls and their (individual and joint) effects on the statistical properties of the collection of (random) per-user SEs. 

These statistical properties include $i)$ the mean and variance of the sum-user SE for finite $m$ and asymptotic $m \to \infty$, $ii)$ the mean and variance of the typical-user SE, again for both finite $m$ and $m \to \infty$, and $iii)$ the asymptotic Chiu-Jain fairness of the collection of per-user SEs.  A key contribution is the explicit expressions for these quantities, given in \thmref{meanstddev} and \thmref{fairness}.  As shown by our numerical investigations for two small networks in \secref{twobslinear} and \secref{quincunx}, the mapping from the bias and power controls to the various performance metrics of interest is non-trivial.  These numerical investigations further demonstrate the significant performance benefits achievable by using both bias and power controls over using either one alone.  The key takeaway is the need for the operator to carefully investigate the often subtle SE performance implications of jointly controlling bias and power in cellular networks.

\subsection{Related Work}
\label{sec:related-work}

The literature on optimal control of both cell biasing and downlink transmission power in cellular networks is vast, and due to space constraints we confine our discussion to the following handful of references, presented more or less chronologically, that are in our opinion most pertinent to our particular model.  Bejerano et al.\ \cite{BejHanLi2007} seek max-min fair associations and show the problem of finding them is NP-hard; they offer a family of association control algorithms for achieving load balancing.  A later paper \cite{BejHan2009} gives optimal algorithms for finding min-max load balancing associations.  Sang et al.\ \cite{SanWanMad2008} introduce the ``weighted $\alpha$-rule'' for opportunistic scheduling of transmissions, where a central server employs cell breathing techniques to balance load across cells.  Son et al.\ \cite{SonChoDeV2009} jointly optimize both partial frequency reuse and load balancing schemes through a network utility maximization (NUM) framework, and through their analysis obtain optimal offline and practical online algorithms. Jo et al.\ \cite{JoSanXia2012} leverage tools from stochastic geometry to derive the downlink SINR cumulative distribution function, the average ergodic rate of the typical user, and the minimum average user throughput, for a $k$-tier HetNet.  Kim et al.\ \cite{KimDeVYan2012} use a NUM framework for user association for flow-level cell load balancing under spatially inhomogeneous traffic, yielding a distributed user association policy that converges to a globally optimal allocation.  Ye et al.\ \cite{YeRonChe2013} study the joint cell association and resource allocation optimization problem.  They decouple the problem into a convex optimization problem, and then develop a distributed algorithm via dual-decomposition, guaranteed to converge with a bounded gap to optimality.  Although each of these references, and many others besides these, deals with cell biasing, to the best of our knowledge none of them explicitly address the question central to our effort: {\em what is the joint impact of both cell biasing and transmit power controls on the SEs achieved by the users?}  

%=========================================================================================
\subsection{Contributions and outline}

\secref{model} introduces our model of a network arena $\Cmc$ holding $n$ BSs, each characterized by location $y_i$, power $t_i$, and bias $b_i$.  These, along with the biased SINR association rule, collectively partition the arena into $n$ cells $\{C_i\}$, one for each BS.  When each BS timeshares uniformly across its users, the powers and cells determine the SE at each possible location.  

\secref{metrics} introduces probability into the model by assuming the $m$ users are positioned independently and uniformly at random over the network arena, naturally leading to the multinomial random vector $\Mbf^{(m)}$ giving the count $M_i$ of users in each cell $\Cmc_i$ with occupancy probability $p_i = |\Cmc_i|/|\Cmc|$.  

After defining the mean and standard deviation of the sum-user and typical-user SE, for  finite $m$ and asymptotic $m \to \infty$, in \defref{metric}, our first key result, \thmref{meanstddev}, gives expressions for these eight quantities in terms of the occupancy probabilities $\{p_i\}$ and the first and second moments of the random instantaneous rate in each cell, $(\phi_i^{(1)},\phi_i^{(2)})$.  \corref{meanstddev} shows the sum-user SE converges in probability to the sum of the mean instantaneous rates within each cell, and the typical user SE converges in probability to zero.  \defref{markowitzbullet} defines the ``Markowitz bullets'' of achievable mean and standard deviation pairs for sum-user ($\Mmc$) and typical-user ($\Mmc_u$) SEs, as the image of these metrics over the set of permissible controls.  

\defref{ChiuJaindef} defines the Chiu-Jain fairness of the set of per-user SEs, and our second key result, \thmref{fairness}, shows fairness converges in probability in $m$ to a constant $\bar{c}$ that is a function of the cell association probabilities and the first two moments of the per-cell instantaneous rates.  \defref{fairnessbullet} defines the throughput-fairness bullet ($\Fmc$) of achievable asymptotic sum-user SE and fairness pairs, as the image of these metrics over the set of permissible controls.  \defref{efficiency} defines the Pareto efficient frontier and Pareto efficient control for all three bullets ($\Mmc$, $\Mmc_u$, $\Fmc$).

\secref{twobslinear} studies the simplest possible non-trivial network: a linear network with two BSs.  We give an explicit expression for the cell boundaries and cell sizes as a function of the control parameters $(\tau,\beta)$ in \prpref{2BScellbdryUnbd}, \corref{2BScellbdryBd}, and \corref{2BScellsize}.  \figref{2BSBoundaries} illustrates these boundaries and cell sizes, and \figref{2BSresults} shows $\Mmc$, $\Mmc_u$, $\Fmc$, and the Pareto efficient frontiers and controls.  The sensitivity of the results to the pathloss function, the number of users, and the arena ``asymmetry'' are discussed.

\secref{quincunx} studies a square arena with five BSs arranged in a quincunx \Cube{5}.  The five cells under various controls are shown in \figref{5bsregions}, and \figref{5bsresults} shows the three bullets, the efficient frontiers, and the efficient controls, as in \figref{2BSresults}.  A key finding is an observed inverse relation between efficient bias and power.

A brief summary is offered in \secref{conclusions}, and the proofs of \thmref{meanstddev} and \thmref{fairness} are found in the Appendix. 

%=========================================================================================
\section{Model}
\label{sec:model}

We consider a bounded arena $\Cmc \subset \Rbb^d$, for $d \geq 1$ the network spatial dimension, containing $n$ fixed BSs, labeled $i \in [n] = \{1,\dots,n\}$, as well as all the mobile users. For each station $i \in [n]$, let $y_i \in \Cmc$ be its location within the arena, and let $t_i \in \Rbb^{+}$ be its assigned downlink transmit power; the corresponding vectors are denoted $\ybf = (y_1,\ldots,y_n)$ and $\tbf = (t_1,\ldots,t_n)$.  All stations employ a common channel for downlink transmissions, with common background noise power $\eta \geq 0$.  Downlink signal power attenuation from each station is subject to a general, but deterministic, pathloss function $l(\cdot,\cdot):\Cmc \times \Cmc \rightarrow \Rbb_+$.  We do not incorporate fading or shadowing.  The signal to interference plus noise ratio (SINR) measured at location $y \in \Cmc$ from the station at $y_i$ is:
\begin{equation}
\label{eq:sinr}
\sinro{y_i,y} = \frac{t_i l(y_i,y)}{\sum_{j \in [n]\setminus i}t_j l(y_j,y) + \eta}.
\end{equation}
Let $(\Cmc_i, i \in [n])$ be a partition of $\Cmc$, where $\Cmc_i$ represents the association region, or cell, for BS $i$, meaning any user at a location $y \in \Cmc_i$ will associate (exclusively) with BS $i$ for downlink transmission.  The cell partition is determined using biased SINR: cell $\Cmc_i$ consists of all locations $y \in \Cmc$ for which the biased SINR from BS $i$ exceeds the biased SINR from all other BSs $j \neq i$:
\begin{equation}
\label{eq:assoc-zone}
\Cmc_i \equiv \left\{y \in \Cmc : b_i \sinro{y_i,y} > b_j \sinro{y_j,y}, j \neq i \right\}.
\end{equation}
The bias $\bbf = (b_1,\ldots,b_n) \in \Rbb_+^n$ and the power $\tbf$ are the two key control knobs studied in this paper.  

Let $m$ be a given number of mobile users, labeled with indices $u \in [m]$, and suppose at some snapshot in time their locations are given by the vector $\zbf = (z_1,\ldots,z_m)$.  The cells $(\Cmc_i, i \in [n])$ and the user locations $\zbf$ enforce the BS-user association, represented by the sets $(\Umc_i, i \in [n])$, with $\Umc_i = \{ u \in [m] : y_u \in \Cmc_i\}$, which together partition $[m]$.  Let $\mbf = (m_1,\ldots,m_n)$ denote the cell occupancies, with $m_i = |\Umc_i|$ and $m_1 + \cdots + m_n = m$.  It will also be useful to define $\Amc(\mbf) = \{i \in [n] : m_i > 0\}$ as the set of occupied cells, and the mapping $i : [m] \to [n]$, with $i(u)$ denoting the BS assigned to user $u$.  

We assume that each BS timeshares its downlink transmissions uniformly among its associated users, with user $u \in \Umc_i$ receiving information $1/m_i$ of the time.  Each user will therefore receive a downlink spectral efficiency of
\begin{equation}
\label{eq:sedef}
x_u^{(m)} = \frac{1}{m_{i(u)}} \log_2 \left(1 + \sinro{y_{i(u)},z_u}\right), ~ u \in [m],
\end{equation}
in units of bits per second per Hertz (bps/Hz), where the superscript $(m)$ emphasizes our interest in understanding the impact of the user population size $m$.  Here, \eqref{sedef} is the Shannon maximum achievable SE over a channel with additive white Gaussian noise (AWGN), treating interference as noise, without fading or shadowing.  Likewise, the overall network spectral efficiency is
\begin{equation}
x^{(m)} = \sum_{u=1}^m x_u^{(m)} = \sum_{i=1}^n \sum_{u \in \Umc_i} x_u^{(m)}.
\end{equation}
\tabref{notation} lists the notation used in the paper.

\begin{table}
\caption{Notation}
\label{tab:notation}
\centering
\renewcommand{\arraystretch}{1.3}
\begin{tabular}{ll} \hline \hline
$\Cmc \subset \Rbb^d$ & network arena \\
$\ybf = (y_1,\ldots,y_n)$ & locations of $n$ BSs \\
$\tbf = (t_1,\ldots,t_n)$ & downlink transmit power \\
$\eta$ & background noise power \\
$l(\cdot,\cdot)$ & general pathloss function \\
$\sinro{y_i,y}$ & SINR at $y$ from BS at $y_i$ \\
$(\Cmc_1,\ldots,\Cmc_n)$ & partition of $\Cmc$ into cells \\
$\bbf = (b_1,\ldots,b_n)$ & cell bias parameters \\
$\zbf = (z_1,\ldots,z_m)$ & locations of mobile users \\
$(\Umc_1,\ldots,\Umc_n)$ & partition of $m$ users into cells \\
$\mbf = (m_1,\ldots,m_n)$ & occupancy of each cell \\
$\Amc(\mbf) \subseteq [n] $ & set of occupied cells \\
$i : [m] \to [n]$ & $i(u)$ is the cell occupied by user $u$ \\
$x_u^{(m)}$ & spectral efficiency (SE) for user $u$ \\
$x^{(m)}$ & overall network SE, sum of $x_u^{(m)}$ \\ \hline
$\Zbf^{(m)}$ & random user locations, uniform iid in $\Cmc$ \\
$\pbf = (p_1,\ldots,p_n)$ & cell association probabilities \\
$\Xbf^{(m)}$ & random user SEs \\
$X^{(m)}$ & random total SE \\
$X_U^{(m)}$ & random SE of randomly selected user $U \in [m]$ \\
$\mu^{(m)},\bar{\mu}$ & mean of total SE \\
$\sigma^{(m)},\bar{\sigma}$ & standard deviation of total SE \\
$\mu^{(m)}_u,\bar{\mu}_u$ & mean of typical user SE \\
$\sigma^{(m)}_u,\bar{\sigma}_u$ & standard deviation of typical user SE \\
$\Mbf^{(m)}$ & random cell occupancy counts \\
$\psi^{(k)} \!\!=\!\! (\psi^{(k)}_i, i \in [n])$ & $k^{\rm th}$ moment of instantaneous transmission rates \\
$\Tmc$ & set of permissible transmission power vectors, $\tbf$ \\
$\Bmc$ & set of permissible cell bias vectors, $\bbf$ \\
$\Mmc^{(m)}$ & SE Markowitz bullets for total SE \\
$\Mmc^{(m)}_u$ & SE Markowitz bullets for typical user SE \\ 
$c(\xbf)$ & Chiu-Jain fairness of the vector $\xbf$ \\
$\bar{c}$ & limiting constant fairness of $c(\Xbf^{(m)})$ \\
$\Fmc$ & SE asymptotic total throughput-fairness bullet \\
$\Emc_{\Mmc},\Emc_{\Mmc_u},\Emc_{\Fmc}$ & efficient controls \\ \hline
$s$ & distance (in meters) separating two BSs \\
$\tau = \log(t_1/t_2)$ & transmit power parameter \\
$\beta = \log(b_1/b_2)$ & cell bias parameter \\
$\Cmc_0$ & cell boundary \\
$\sigma(\tau,\beta,\alpha)$ & cell boundary parameter \\
$\gamma(\sigma)$ & cell boundary function \\
$\tilde{y}_{\pm}(\gamma)$ & cell boundary function \\
$\Cmc = [-\Delta_1,+\Delta_2]$ & network arena interval \\
$h(\Delta)$ & parameter $\sigma$ for which $\tilde{y}_{\pm}$ hits $\Delta_{1/2}$ \\
$\delta$ & ``dead-zone'' reception radius around transmitter \\ \hline
\end{tabular}
\end{table}

%=========================================================================================
\section{Metrics}
\label{sec:metrics}

We henceforth assume the user locations to be i.i.d. $\Cmc$-valued random variables, denoted $\Zbf^{(m)} = (Z_1,\ldots,Z_m)$, each uniformly distributed over $\Cmc$.  Thus $\Pbb(Z^{(m)}_u \in \Cmc') = |\Cmc'|/|\Cmc|$, for each $\Cmc' \subseteq \Cmc$, and in particular, the induced cell association probabilities are given by $\pbf = (p_1,\ldots,p_n)$, with 
\begin{equation}
\label{eq:assocprob}
p_i = \frac{|\Cmc_i|}{|\Cmc|}, ~i \in [n],
\end{equation}
where $p_1 + \cdots + p_n = 1$.  Let $\bar{p}$ denote $1-p$.  The random locations induce random user downlink SEs, denoted $\Xbf^{(m)} = (X_1^{(m)},\ldots,X_m^{(m)})$, and a random total (sum-user) downlink SE, denoted $X^{(m)}$.  The random SE for a ``typical'' user is the RV $X^{(m)}_U$, for $U \in [m]$ selected uniformly at random.  

The random locations $\Zbf^{(m)}$ induce a random partition of cell associations $(\Umc_1,\ldots,\Umc_n)$ among the MUs.  This random partition in turn induces a random vector of cell occupancy counts $\Mbf^{(m)} = (M^{(m)}_1,\ldots,M^{(m)}_n)$.  Here, $M^{(m)}_i = |\Umc_i|$ for $i \in [n]$ is a binomial random variable with parameters $m$ and $p_i$, denoted $M^{(m)}_i \sim \mathrm{bin}(m,p_i)$.  The random vector $\Mbf^{(m)}$ has a multinomial distribution, denoted $\Mbf^{(m)} \sim \mathrm{mult}(m,\pbf)$, support $\Mmc^{(m)} = \{ \mbf = (m_1,\ldots,m_n) \in \Nbb^n : m_1+\cdots+m_n = m\}$, and, for $\binom{m}{\mbf}$ the multinomial coefficient:
\begin{equation}
\label{eq:multdisbn}
\Pbb(\Mbf^{(m)} = \mbf) = \binom{m}{\mbf} \prod_{i=1}^n p_i^{m_i}. 
\end{equation}

In what follows it will be useful to define the vectors $\psi = (\psi_i, i \in [n])$ and $\psi^{(2)} = (\psi^{(2)}_i, i \in [n])$, with 
\begin{equation}
\psi^{(k)}_i = \frac{1}{|\Cmc_i|} \int_{\Cmc_i} (\log_2(1+\sinro{y_i,y}))^k \drm y, ~ k \in \Nbb.
\label{eq:expectinstrateint}
\end{equation}
When $k=1$ we drop the superscript and write $\psi^{(k)}_i = \psi_i$, and emphasize $\psi^{(2)}_i \neq \psi_i^2 = (\psi^{(1)}_i)^2$.  Here, $\psi^{(k)}_i$ represents the $k^{\rm th}$ moment of the cell $i$ instantaneous transmission rate, respectively, where the expectation is with respect to the random location of the user being uniformly distributed in $\Cmc_i$. It follows that $\psi^{(2)}_i - \psi_i^2$ is the variance of the instantaneous transmission rate in cell $i$. Note  $\psi^{(k)}_i$ is independent of $m$.

\subsection{Mean and standard deviation of SE}
\label{ssec:model-meanstddev}

In this subsection we derive expressions for four metrics pertaining to downlink SE in the model described in \secref{model}: the mean and standard deviation of the total (sum over all users) SE, and the mean and standard deviation of the typical user SE.  For each of the four metrics we obtain expressions for both a finite $m$ number of users and the limit as $m \to \infty$.
\begin{definition}
\label{def:metric}
The mean and standard deviation of the total and typical user SE, for both finite and infinite $m$, are defined:
\begin{equation}
\begin{array}{cclccl}
\mu^{(m)} &=& \Ebb[X^{(m)}] & \sigma^{(m)} &=& \mathrm{Std}(X^{(m)}) \\
\bar{\mu} &=& \lim_{m \to \infty} \mu^{(m)} & \bar{\sigma} &=& \lim_{m \to \infty} \sigma^{(m)} \\
\mu^{(m)}_u &=& \Ebb[X_U^{(m)}] & \sigma^{(m)}_u &=& \mathrm{Std}(X_U^{(m)}) \\
\bar{\mu}_u &=& \lim_{m \to \infty} \mu^{(m)}_u & \bar{\sigma}_u &=& \lim_{m \to \infty} \sigma^{(m)}_u 
\end{array}
\end{equation}
\end{definition}

\begin{theorem}
\label{thm:meanstddev}
When users are distributed uniformly at random, the eight SE metrics defined in \defref{metric} are as follows.  First:
\begin{equation}
\mu^{(m)} = \sum_{i=1}^n (1-\bar{p}_i^m) \psi_i.
\end{equation}
and $\bar{\mu} = \sum_{i=1}^n \psi_i$.  Second:
\begin{eqnarray}
\label{eq:varm}
\left(\sigma^{(m)}\right)^2 
&=& \sum_{i=1}^n (\psi^{(2)}_i - \psi_i^2) \Ebb\left[\frac{1}{M_i^{(m)}} \mathbf{1}_{M_i^{(m)} > 0} \right] \\
&+& \sum_{i=1}^n \psi_i^2 (1-\bar{p}_i^m)\bar{p}_i^m \nonumber \\
&+& 2 \sum_{1 \leq i < j \leq n} \psi_i \psi_j \left( (1-(p_i+p_j))^m - (\bar{p}_i \bar{p}_j)^m \right), \nonumber 
\end{eqnarray}
and $\bar{\sigma}^2 = 0$.  Third:
\begin{equation}
\mu_u^{(m)} = \frac{\mu^{(m)}}{m}.
\end{equation}
and $\bar{\mu}_u = 0$.  Fourth:
\begin{eqnarray}
\label{eq:sigum}
\left(\sigma_u^{(m)}\right)^2 
&=& \frac{1}{m} \sum_{i=1}^n \psi^{(2)}_i \Ebb\left[ \frac{1}{M_i^{(m)}} \mathbf{1}_{M_i^{(m)} > 0} \right] \\
& - & \frac{1}{m^2} \sum_{i=1}^n \psi_i^2 (1-\bar{p}_i^m)^2 \nonumber \\
& - & \frac{2}{m^2} \sum_{1 \leq i < j \leq n} \psi_i \psi_j \left(  1 + (\bar{p}_i\bar{p}_j)^m - \bar{p}_i^m - \bar{p}_j^m \right), \nonumber 
\end{eqnarray}
and $\bar{\sigma}_u^2 = 0$.
\end{theorem}
The proof is given in the appendix.  Recall that Chebychev's inequality applied to the definition of convergence in probabality, denoted $\stackrel{p}{\to}$, ensures that if $\{V^{(m)}\}$ is a sequence of RVs with $\Ebb[V^{(m)}] \to \nu$ and $\mathrm{Var}(V^{(m)}) \to 0$ then $V^{(m)} \stackrel{p}{\to} \nu$.  
\begin{corollary}
\label{cor:meanstddev}
The total SE and typical user SE converge in probability to constants:
\begin{equation}
\sum_{u=1}^m X^{(m)}_u \stackrel{p}{\to} \sum_{i=1}^n \psi_i, ~ X^{(m)}_U \stackrel{p}{\to} 0.
\end{equation}
\end{corollary}

\begin{figure}[!ht]
\centering
\includegraphics[width=3.5in]{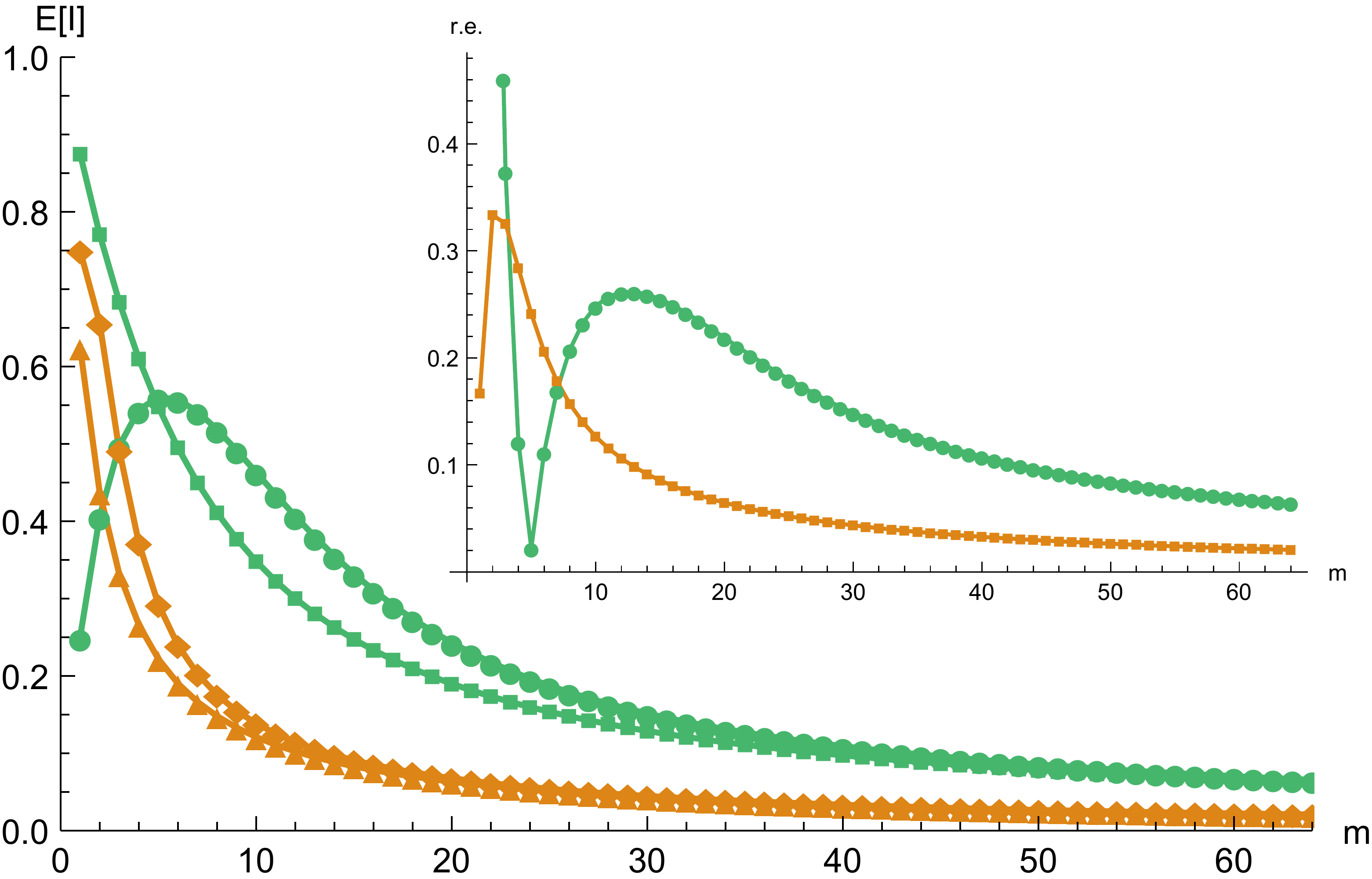}
\caption{$\Ebb[I^{(m)}]$ and its approximation $\Ebb[1/(M^{(m)}+1)]$ vs.\ $m$ for $m=1/4$ (green) and $m=3/4$ (tan); both go to $0$ in $m$ as $1/m$.  Inset: relative error of the approximation vs.\ $m$.}
\label{fig:exp1mandre}
\end{figure}

Computing $\sigma^{(m)}$ and $\sigma^{(m)}_u$ in \thmref{meanstddev} requires computing $\Ebb[I^{(m)}]$ for $I^{(m)} = \frac{1}{M^{(m)}}\mathbf{1}_{M^{(m)}>0}$ and $M^{(m)} \sim \mathrm{bin}(m,p)$.  As the support of $M^{(m)}$ is $\{0,\ldots,m\}$, the running time of this computation is $O(m)$, which may be significant for sufficiently large $m$.  Although we have established $\Ebb[I^{(m)}] \to 0$ as $m \to \infty$ in \lemref{exp1m}, it is still useful to have an approximation for this sum with constant running time.  A natural choice  is
\begin{equation}
\Ebb \left[ \frac{1}{M^{(m)}+1} \right] = \frac{1-\bar{p}^{m+1}}{p(m+1)}.
\end{equation}
\figref{exp1mandre} shows the quantities $\Ebb[I^{(m)}]$ and $\Ebb[1/(M^{(m)}+1)]$ vs.\ $m$.  Note $\Ebb[I^{(m)}] > \Ebb[1/(M^{(m)}+1)]$ for $m > m(p)$.

%The asymptotic SEs $(\bar{\mu},\bar{\sigma},\bar{\mu}_u,\bar{\sigma}_u)$ are all zero, aside from $\bar{\mu}=\sum_i \psi_i$.  Thus, for large $m$ the total SE is $\bar{\mu}$ with high probability, and each of the individual user SEs are negligibly small.  Physically, such a network is overloaded, with each user receiving unsatisfactory rate, and so it is of more interest to focus on the finite $m$ regime instead of the asymptotic regime.  

In the finite $m$ regime one might intuit that there is a natural tradeoff between the SE mean and standard deviation, both for total and typical user SE.  To investigate this hypothesis, we borrow the concept of the Markowitz bullet from the portfolio optimization problem in finance.  In this problem the return of each possible portfolio of risky and riskless assets has an associated mean $\mu$ and standard deviation $\sigma$, and the set of achievable $(\sigma,\mu)$ pairs (shown on the $\sigma-\mu$ plane), as one sweeps over all possible portfolios, forms the Markowitz ``bullet'', so-called because of its shape.  %The efficient (or Pareto) frontier of this bullet is the set of non-dominated  $(\sigma,\mu)$ pairs, meaning there does not exist any feasible point in the bullet with both smaller $\sigma$ and larger $\mu$.  This frontier is the market tradeoff between ``risk'' ($\sigma$) and ``reward'' ($\mu$).  

In our context, the two controls of interest are the downlink transmission power vector $\pbf$ and the bias vector $\bbf$.  A key focus of this paper is to understand the types of ``risk-reward'' tradeoffs achievable with both unilateral and bilateral control, where unilateral control refers to optimizing one of $\tbf,\bbf$ holding the other fixed, and bilateral control refers to jointly optimizing $\tbf$ and $\bbf$.  Let $\Tmc$ be the set of permissible transmission power vectors $\tbf$, and let $\Bmc$ be the set of permissible bias vectors $\bbf$.  
\begin{definition}
\label{def:markowitzbullet}
The total SE bilateral control Markowitz bullet is:
\begin{equation}
\Mmc^{(m)}(\Tmc,\Bmc) = \{ (\sigma^{(m)}(\tbf,\bbf),\mu^{(m)}(\tbf,\bbf)) : (\tbf,\bbf) \in \Tmc \times \Bmc \} 
\end{equation}
with unilateral control bullets $\Mmc^{(m)}(\Tmc,\mathbf{1}),\Mmc^{(m)}(\mathbf{1},\Bmc)$.  The typical user SE bilateral control Markowitz bullet is:
\begin{equation}
\Mmc^{(m)}_u(\Tmc,\Bmc) = \{ (\sigma^{(m)}_u(\tbf,\bbf),\mu^{(m)}_u(\tbf,\bbf)) : (\tbf,\bbf) \in \Tmc \times \Bmc\} 
\end{equation}
with unilateral control bullets $\Mmc^{(m)}_u(\Tmc,\mathbf{1}),\Mmc^{(m)}_u(\mathbf{1},\Bmc)$.  
\end{definition}
Assuming $\tbf = \mathbf{1}$, instead of some constant scaling, say $\tbf = \rho \mathbf{1}$ for $\rho > 0$, as a default for transmission powers is natural in the low noise regime, where the SINR is approximately equal to the SIR.  The SIR is homogeneous of degree $0$, i.e., $\mathrm{sir}(\rho \tbf) = \mathrm{sir}(\tbf)$, meaning scaling by $\rho$ has no effect on the SIR, and thus it is natural to choose $\rho = 1$.

\begin{figure}[!ht]
\centering
\includegraphics[width=3.5in]{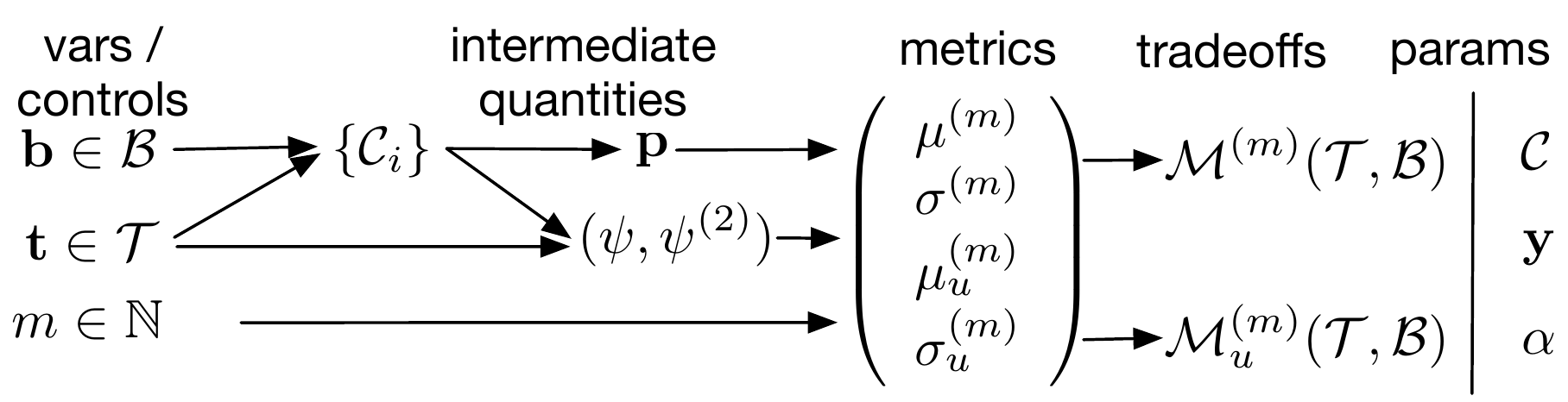}
\caption{Relationships among key quantities: $i)$ variables / controls include cell bias $\bbf \in \Bmc$, transmit power $\tbf \in \Tmc$, and number of users $m \in \Nbb$; $ii)$ intermediate quantities include cells $\{\Cmc_i\}$ \eqref{assoc-zone}, association probabilities $\pbf$ \eqref{assocprob}, and expected instantaneous rate vectors $(\psi,\psi^{(2)})$ \eqref{expectinstrateint}; $iii)$ performance metrics include means and standard deviations for  sum and typical user SE (\defref{metric} and \thmref{meanstddev}); $iv)$ tradeoffs of metrics are captured by Markowitz bullets (\defref{markowitzbullet}); $v)$ ``given'' parameters include the  arena $\Cmc$, BS locations $\ybf$, and pathloss exponent $\alpha$.}
\label{fig:variablediagram}
\end{figure}

Observe that the controls $(\tbf,\bbf)$ relate to the metrics $(\sigma^{(m)},\mu^{(m)},\sigma^{(m)}_u,\mu^{(m)}_u)$ through the intermediate quantities $(\{\Cmc_i\},\pbf,\psi,\psi^{(2)})$, illustrated in \figref{variablediagram}.  Observe that the metrics are functions of the inputs $\pbf,\psi,\psi^{(2)}$ (which depend upon the controls $\bbf,\tbf$ but not on $m$), and the variable $m$.  Alg.~\ref{alg:markowitzbullet} exploits these relationships for computing $\Mmc^{(m)}(\Tmc,\Bmc)$ by first computing and storing an association (hash) $(\tbf,\bbf) \to (\pbf,\psi,\psi^{(2)})$ for each $(\tbf,\bbf) \in \Tmc \times \Bmc$ (independent of $m$), and then using this map, along with $m$, to find the association (hash) $(\tbf,\bbf,m) \to (\sigma^{(m)},\mu^{(m)})$, which enables plotting $\Mmc^{(m)}(\Tmc,\Bmc)$.  The algorithm for $\Mmc_u^{(m)}$ is analogous.  The computational difficulty arises from choosing a suitably fine bounding of and discretization for the control space (e.g, $\Tmc \times \Bmc$) and a suitably fine discretization of the arena $\Cmc$.  

\begin{algorithm}[t!]
  \caption{SE Markowitz bullet $\Mmc^{(m)}(\Tmc,\Bmc)$ algorithm}
  \label{alg:markowitzbullet}
  \begin{algorithmic}[1]
  \State {\bf input:} arena $\Cmc$, BS locations $\ybf$, pathloss $\alpha$, controls $(\Tmc,\Bmc)$
  \ForAll{feasible controls $(\tbf,\bbf) \in \Tmc \times \Bmc$} 
  	\State {\bf initialize:} cells $\Cmc_i = \emptyset$ for $i \in [n]$
  	\ForAll{locations $y \in \Cmc$}
  		\State determine $y$'s association $i \in [n]$, add to $\Cmc_i$
  	\EndFor
	\State compute $\pbf$ from $\{\Cmc_i\}$ \& $(\psi,\psi^{(2)})$ from $(\{\Cmc_i\},\tbf)$
  \EndFor
  \State {\bf store:} association map $(\tbf,\bbf) \to (\pbf,\psi,\psi^{(2)})$
  \State 
  \State {\bf input:} above association map and number of users $m$
  \State {\bf initialize:} $\Mmc^{(m)}(\Tmc,\Bmc) = \emptyset$
  \ForAll{feasible controls $(\tbf,\bbf)\in \Tmc \times \Bmc$} 
  	\State lookup $(\pbf,\psi,\psi^{(2)})$ for current $(\tbf,\bbf)$
  	\State compute $(\sigma^{(m)},\mu^{(m)})$ from $m$ and $(\pbf,\psi,\psi^{(2)})$
	\State add $(\tbf,\bbf,m) \to (\sigma^{(m)},\mu^{(m)})$ to $\Mmc^{(m)}(\Tmc,\Bmc)$
  \EndFor
  \State {\bf return:} $\Mmc^{(m)}(\Tmc,\Bmc)$
  \end{algorithmic}
\end{algorithm}

\subsection{Chiu-Jain fairness of SE}

% SW: add motivation: since variance goes to 0 in m, there is no asymptotic mean-variance tradeoff, and the non-asymptotic expressions are difficult to work with.  The variance is not the best measure of inequality across users, since it goes to zero in m, even though for any given m there may be substantial variation across users.

We next study the asymptotic Chiu-Jain fairness \cite{JaiChiHaw1984} (hereafter, simply fairness) of the vector of per-user random SEs $\Xbf^{(m)} = (X^{(m)}_1,\ldots,X^{(m)}_m)$.  The fairness of an $m$-vector $\xbf$ is:
\begin{equation}
\label{eq:ChiuJaindef}
c(\xbf)=\frac{\left(\sum_{u=1}^m x_u\right)^2}{m\sum_{u=1}^m x_u^2} \in \left[\frac{1}{m},1\right].
\end{equation}
Here $c(\xbf) = 1/m$ for $\xbf$ any unit vector $(\ebf_1,\ldots,\ebf_m)$, and $c(\xbf)=1$ for any constant vector. 
\begin{definition}
\label{def:ChiuJaindef}
The fairness of the random SEs $\Xbf^{(m)} = (X^{(m)}_1,\ldots,X^{(m)}_m)$ is the random variable $c(\Xbf^{(m)})$.  
\end{definition} 

\begin{theorem}
\label{thm:fairness}
The (random) fairness of the random vector of per-user SEs $c(\Xbf^{(m)})$ converges in probability to a constant:
\begin{equation}
\label{eq:chiujainasym}
c(\Xbf^{(m)}) \stackrel{p}{\to} \bar{c} = \frac{\left(\sum_{i=1}^n \psi_i \right)^2}{\sum_{i=1}^n \frac{\psi^{(2)}_i}{p_i}} \mbox{ as } m \to \infty.
\end{equation}
\end{theorem}
The proof is given in the appendix.

This theorem motivates our second key performance tradeoff: asymptotic total SE throughput vs.\ fairness.
\begin{definition}
\label{def:fairnessbullet}
The asymptotic (in $m$) total SE bilateral control throughput fairness bullet is defined as:
\begin{equation}
\Fmc(\Tmc,\Bmc) = \{ (\bar{c}(\tbf,\bbf),\bar{\mu}(\tbf,\bbf)) : (\tbf,\bbf) \in \Tmc \times \Bmc \} 
\end{equation}
for $\bar{c}$ in \thmref{fairness} and $\bar{\mu}$ in \thmref{meanstddev}, with the unilateral control bullets given by $\Fmc(\Tmc,\mathbf{1})$ and $\Fmc(\mathbf{1},\Bmc)$.  
\end{definition}
The following definition establishes the notion of efficiency.
\begin{definition}
\label{def:efficiency}
For each tradeoff, $(\Mmc^{(m)},\Mmc^{(m)}_u,\Fmc)$, the {\em (Pareto) efficient frontier} for a given $(\Tmc,\Bmc)$ and $m$ is the corresponding subset of non-dominated achievable points:
\begin{eqnarray}
\hat{\Mmc}^{(m)} \!\!\!\!\! &=& \!\!\!\!\! \{ \xi \in \Mmc^{(m)} : \not\exists \xi' \in \Mmc^{(m)} : \sigma' < \sigma \& \mu' > \mu\} \nonumber \\
\hat{\Mmc}^{(m)}_u \!\!\!\!\! &=& \!\!\!\!\! \{ \xi_u \in \Mmc^{(m)}_u : \not\exists \xi_u' \in \Mmc^{(m)}_u : \sigma_u' < \sigma_u \& \mu_u' > \mu_u \} \nonumber \\
\hat{\Fmc} \!\!\!\!\! &=& \!\!\!\!\! \{ \bar{\xi} \in \Fmc : \not\exists \bar{\xi}' \in \Fmc : \bar{c}' > \bar{c} \; \& \; \bar{\mu}' > \bar{\mu} \}, 
\end{eqnarray}
where $\xi = (\sigma,\mu)$, $\xi_u = (\sigma_u,\mu_u)$, and $\bar{\xi} = (\bar{c},\bar{\mu})$.  
The {\em (Pareto) efficient controls} are those $(\tbf,\bbf)$ points achieving the corresponding efficient frontiers:
\begin{eqnarray}
\Emc_{\Mmc}^{(m)} &=& \{ (\tbf,\bbf) \in \Tmc \times \Bmc : (\sigma^{(m)},\mu^{(m)}) \in \hat{\Mmc}^{(m)} \} \nonumber \\
\Emc_{\Mmc_u}^{(m)} &=& \{ (\tbf,\bbf) \in \Tmc \times \Bmc : (\sigma^{(m)}_u,\mu^{(m)}_u) \in \hat{\Mmc_u}^{(m)} \} \nonumber \\
\Emc_{\Fmc} &=& \{ (\tbf,\bbf) \in \Tmc \times \Bmc : (\bar{c},\bar{\mu}) \in \hat{\Fmc} \} 
\end{eqnarray}
\end{definition}
Non-dominating inequalities are different for $(\Mmc,\Mmc_u)$ and $\Fmc$, since lower variance and greater fairness are desirable.  

The next two sections illustrate the three key performance tradeoffs $\Mmc^{(m)}(\Tmc,\Bmc)$, $\Mmc^{(m)}_u(\Tmc,\Bmc)$, and $\Fmc(\Tmc,\Bmc)$, along with the unilateral variants $(\Tmc,\mathbf{1})$ and $(\mathbf{1},\Bmc)$, and their Pareto efficient frontiers and controls, in $i)$ a two BS linear network (\secref{twobslinear}), and $ii)$ a five BS ``quincunx'' network (\secref{quincunx}).  

%=========================================================================================
\section{Two BS linear network}
\label{sec:twobslinear}

\subsection{Model}
\begin{figure}[!ht]
\centering
\includegraphics[width=3.0in]{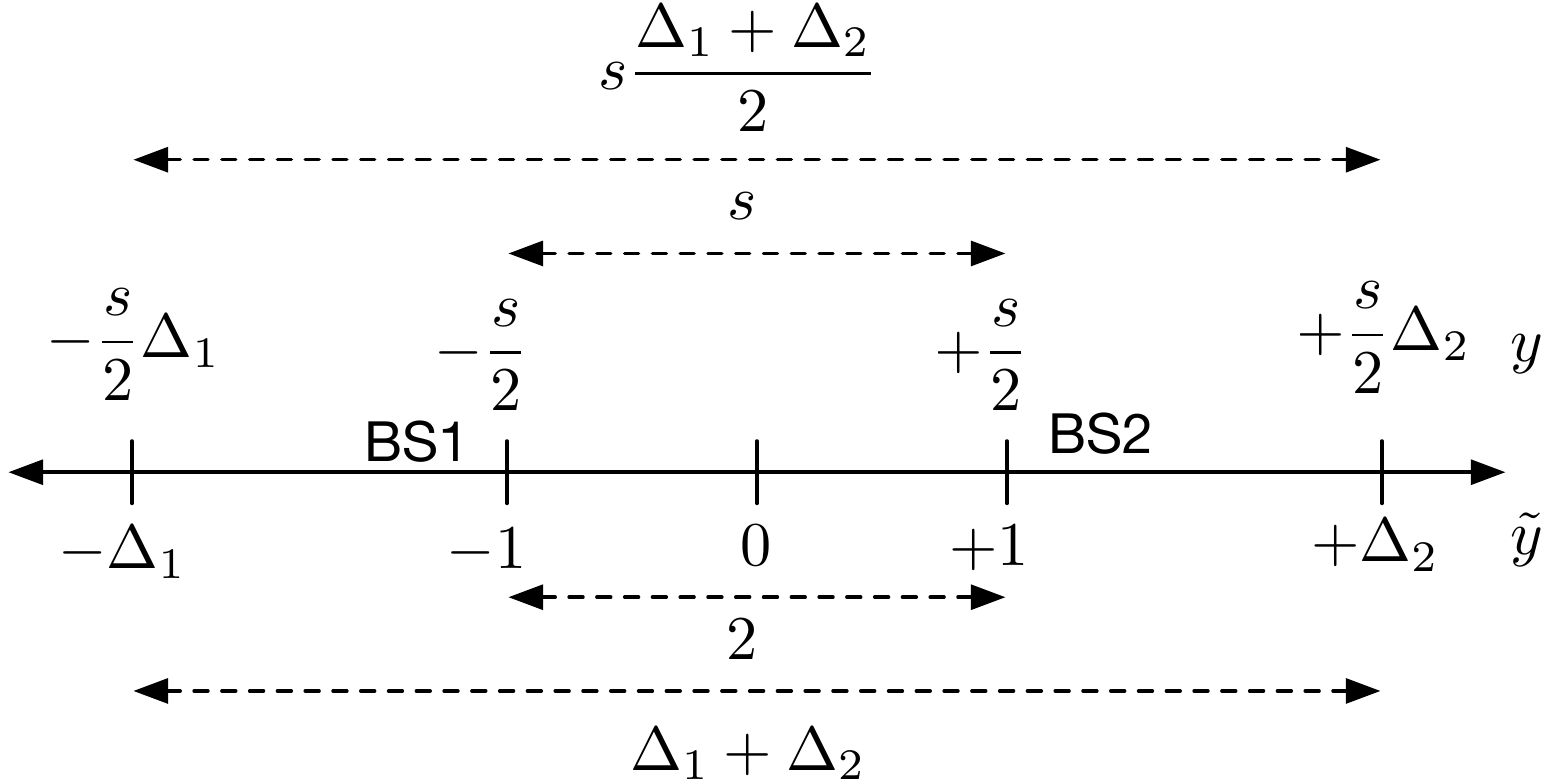}
\caption{(\secref{twobslinear}) Two BS linear network.  Quantities above (below) the line are measured in meters (normalized units of $s/2$ meters), with label $y$ ($\tilde{y}$), respectively.}
\label{fig:2bstopology}
\end{figure}

Consider a cellular ``network'' with two BSs, separated by distance $s > 0$ meters, as shown in \figref{2bstopology}.  Without loss of generality let the two BSs be at positions $y_1 = -s/2$ and $y_2 = +s/2$ on the infinite line connecting their positions.  For simplicity, the network domain $\Cmc$ is restricted to be a segment of this infinite line.  It is convenient to measure space (locations and distances) in units of $s/2$, with location $y \in \Cmc$ at distance $|y|$ meters from the origin, reported at position $\tilde{y} = 2y/s$ at $|\tilde{y}| = 2|y|/s$ lengths from the origin.  We henceforth measure spatial quantities in terms of such lengths, and in particular, the BS locations are at $\tilde{y}_1 = -1$ and $\tilde{y}_2 = +1$.

We use the standard pathloss model for signal attenuation with pathloss exponent $\alpha \geq 1$, i.e., $l(y_1,y_2) = |y_1-y_2|^{-\alpha}$ represents the received signal power at location $y_2$ from a unit power transmitter at location $y_1$.  Using the above change of spatial units, $l(y_1,y_2)$ above becomes $\hat{l}(\tilde{y}_1,\tilde{y}_2) = (2/s)^{\alpha} |\tilde{y}_1-\tilde{y}_2|^{-\alpha}$.  To address the pathloss singularity at the transmitter, we augment this model with a ``dead-zone'' of radius $\delta \geq 0$, so that the pathloss is only defined if $|\tilde{y}_1-\tilde{y}_2| > \delta$, and inside this radius no reception is possible.  For tractability, in this section we drop the noise term ($\eta = 0$), and as such our SINR becomes a signal to interference ratio (SIR).  In particular, the downlink SIR seen at location $\tilde{y} \in \Rbb$ from the two BSs are
\begin{equation}
\siro{\tilde{y}_1,\tilde{y}} = \frac{t_1 \hat{l}(\tilde{y}_1,\tilde{y})}{t_2 \hat{l}(\tilde{y}_2,\tilde{y})}, ~ 
\siro{\tilde{y}_2,\tilde{y}} = \frac{t_2 \hat{l}(\tilde{y}_2,\tilde{y})}{t_1 \hat{l}(\tilde{y}_1,\tilde{y})},
\end{equation}
where $\tbf = (t_1,t_2) \in \Rbb^2_+$ is the downlink transmission power vector.  Define the {\em transmit power parameter}: 
\begin{equation}
\tau = \log \left( \frac{t_1}{t_2} \right) \in \Rbb_+,
\end{equation}
so that 
\begin{equation}
\siro{\tilde{y}_1,\tilde{y}} = \erm^{\tau} \left( \frac{|\tilde{y}-1|}{|\tilde{y}+1|}\right)^{\alpha} \!\!\!\!, ~
\siro{\tilde{y}_2,\tilde{y}} = \erm^{-\tau} \left( \frac{|\tilde{y}+1|}{|\tilde{y}-1|}\right)^{\alpha}\!\!\!\!.
\end{equation}
For the case of two BSs, the bias parameters $\bbf = (b_1,b_2) \in \Rbb^2_+$ determine the cell boundary through their ratio, $b_1/b_2$.  We reparameterize using the {\em cell bias parameter}:
\begin{equation}
\beta = \log \left(\frac{b_1}{b_2} \right) \in \Rbb_+.
\end{equation}  
Define $(\Cmc_0,\Cmc_1,\Cmc_2)$ as a partition of the network domain $\Cmc$: 
\begin{eqnarray}
\Cmc_0 &=& \{ \tilde{y} \in \Cmc : \erm^{\beta} \siro{\tilde{y}_1,\tilde{y}} = \siro{\tilde{y}_2,\tilde{y}} \} \nonumber \\
\Cmc_1 &=& \{ \tilde{y} \in \Cmc : \erm^{\beta} \siro{\tilde{y}_1,\tilde{y}} > \siro{\tilde{y}_2,\tilde{y}} \} \nonumber \\
\Cmc_2 &=& \{ \tilde{y} \in \Cmc : \erm^{\beta} \siro{\tilde{y}_1,\tilde{y}} < \siro{\tilde{y}_2,\tilde{y}} \} \label{eq:2BScellbndrydef}
\end{eqnarray}
We refer to $\Cmc_0$ as the cell boundary.  

\subsection{Cell boundary and cell lengths}

Define the following pair of functions $\sigma$ and $\gamma$:
\begin{equation}
\label{eq:2BScellbdrysigmagamma}
\sigma = \sigma(\tau,\beta,\alpha) = \frac{2\tau+\beta}{\alpha}, ~ \gamma = \gamma(\sigma) = \frac{\erm^{\sigma}+1}{\erm^{\sigma}-1}.
\end{equation}
The following proposition gives $\Cmc_0$ for the case $\Cmc = \Rbb$.   
\begin{proposition}
\label{prp:2BScellbdryUnbd}
For $\Cmc = \Rbb$ with cell bias parameter $\beta$ and transmit power parameter $\tau$, $\Cmc_0 = \{\tilde{y}_-,\tilde{y}_+\}$, with 
\begin{equation}
\tilde{y}_{\pm} = \tilde{y}_{\pm}(\gamma) = \gamma \pm \sqrt{\gamma^2-1},
\end{equation}
for $\gamma = \gamma(\sigma)$ and $\sigma = \sigma(\tau,\beta,\alpha)$ in \eqref{2BScellbdrysigmagamma}.  $(\Cmc_1,\Cmc_2)$ are: 
\begin{equation}
\label{eq:2BScellbdryC1C2}
\begin{array}{c|cc}
\sigma & \Cmc_1 & \Cmc_2 \\ \hline
<0 & (\tilde{y}_{-},\tilde{y}_{+}) & (-\infty,\tilde{y}_{-}) \cup (\tilde{y}_{+},\infty) \\
>0 & (-\infty,\tilde{y}_{-}) \cup (\tilde{y}_{+},\infty) & (\tilde{y}_{-},\tilde{y}_{+}) 
\end{array}
\end{equation}
\end{proposition}
\begin{IEEEproof}
By \eqref{2BScellbndrydef}, a boundary point $\tilde{y}$ must satisfy
\begin{eqnarray}
0 &=& \erm^{\beta} \siro{\tilde{y}_1,\tilde{y}} -\siro{\tilde{y}_2,\tilde{y}} \nonumber \\
0 &=& \erm^{\frac{2\tau+\beta}{\alpha}} \frac{|\tilde{y}-1|}{|\tilde{y}+1|}- \frac{|\tilde{y}+1|}{|\tilde{y}-1|} \nonumber \\
0 &=& \erm^{\sigma} (\tilde{y}-1)^2 - (\tilde{y}+1)^2 \nonumber \\
0 &=& \left(\erm^{\sigma}-1\right) \tilde{y}^2 - 2 \left(\erm^{\sigma}+1\right) \tilde{y} + \left(\erm^{\sigma}-1\right) \nonumber \\
0 &=& \tilde{y}^2 - 2 \gamma \tilde{y} + 1 \nonumber \\
\tilde{y} &=& \gamma \pm \sqrt{ \gamma^2 - 1}
\end{eqnarray}
where the sequence of equations are all equivalent.  
\end{IEEEproof}

\begin{figure*}
\centering
\includegraphics[width=\textwidth]{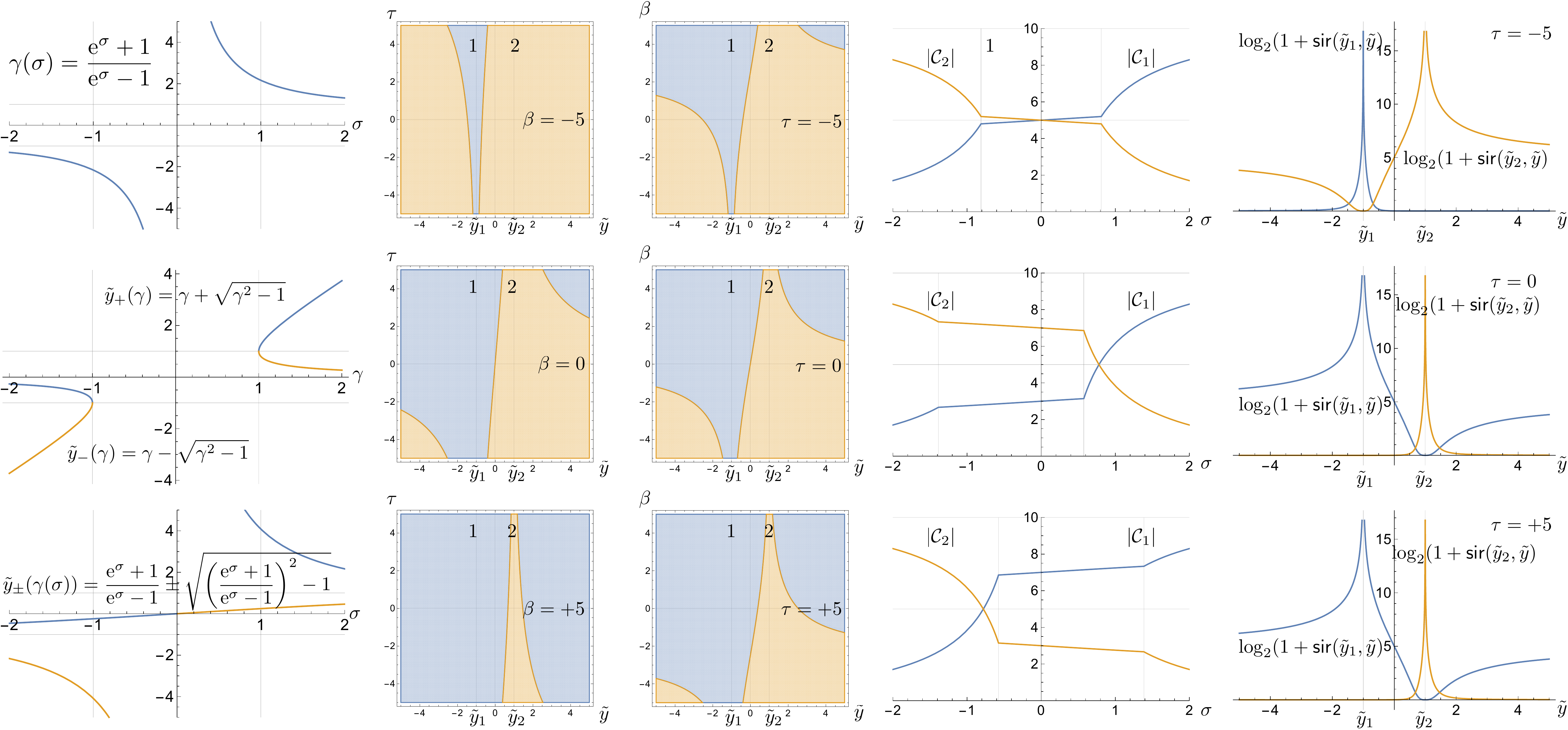}
\caption{(\secref{twobslinear}) The two BS network for $\alpha=3$.  Column 1: $\gamma(\sigma)$, $\tilde{y}_{\pm}(\gamma)$, and $\tilde{y}_{\pm}(\gamma(\sigma))$.  Column 2: $\Cmc_1$ (blue) and $\Cmc_2$ (tan) on horizontal axis vs.\ $\tau$ on vertical axis for $\beta \in \{-5,0,+5\}$.  Column 3: $\Cmc_1$ (blue) and $\Cmc_2$ (tan) on horizontal axis vs.\ $\beta$ on vertical axis for $\tau \in \{-5,0,+5\}$.  Column 4: $|\Cmc_1|$ and $|\Cmc_2|$ vs.\ $\sigma$ for $\Cmc = [-\Delta_1,+\Delta_2] \in \{[-5,+5],[-3,+7],[-7,+3]\}$.  Column 5: $\log_2(1+\siro{\tilde{y}_1,\tilde{y}})$ and $\log_2(1+\siro{\tilde{y}_2,\tilde{y}})$ vs.\ $\tilde{y}$ for $\tau \in \{-5,0,+5\}$.}
\label{fig:2BSBoundaries}
\end{figure*}

The functions $\gamma(\sigma)$, $\tilde{y}_{\pm}(\gamma)$, and $\tilde{y}_{\pm}(\gamma(\sigma))$ are shown in \figref{2BSBoundaries}.  We henceforth restrict the network arena from $\Cmc = \Rbb$ to $\Cmc = [-\Delta_1,+\Delta_2]$ for $\Delta_i > 1$, $i \in \{1,2\}$.  Define 
\begin{equation}
h(\Delta) = 2 \log \left(\frac{\Delta+1}{\Delta-1}\right).
\end{equation}
For $\sigma < 0$ the function $\tilde{y}_{-}(\gamma(\sigma))$ hits $-\Delta_1$ at $\sigma = -h(\Delta_1)$ and the function $\tilde{y}_{+}(\gamma(\sigma))$ hits $+\Delta_2$ at $\sigma = +h(\Delta_2)$.

\begin{corollary}
\label{cor:2BScellbdryBd}
For $\Cmc = [-\Delta_1,+\Delta_2]$ with cell bias parameter $\beta$ and transmit power parameter $\tau$, the cells $(\Cmc_1,\Cmc_2)$ are the following functions of $\tilde{y}_{\pm} = \tilde{y}_{\pm}(\gamma(\sigma))$, for $\sigma = \sigma(\tau,\beta,\alpha)$:
\begin{equation}
\begin{array}{r|cc}
& \Cmc_1 & \Cmc_2 \\ \hline
i) & (\tilde{y}_{-},\tilde{y}_{+}) & (-\Delta_1,\tilde{y}_{-}) \cup (\tilde{y}_{+},+\Delta_2) \nonumber \\
ii) & (-\Delta_1,\tilde{y}_{+}) & (\tilde{y}_{+},+\Delta_2) \\
iii) & (-\Delta_1,0) & (0,+\Delta_2) \\
iiv) & (-\Delta_1,\tilde{y}_{-}) & (\tilde{y}_{-},+\Delta_2) \\
v) & (-\Delta_1,\tilde{y}_{-}) \cup (\tilde{y}_{+},+\Delta_2) & (\tilde{y}_{-},\tilde{y}_{+})
\end{array}
\end{equation}
where the five cases are: $i)$ $\sigma < -h(\Delta_1)$, $ii)$ $-h(\Delta_1) < \sigma < 0$, $iii)$ $\sigma = 0$, $iv)$ $0 < \sigma < +h(\Delta_2)$, and $v)$ $\sigma > +h(\Delta_2)$.
\end{corollary}

The cells $\Cmc_1,\Cmc_2$ as functions of $\tau$ and $\beta$ are shown in \figref{2BSBoundaries}.  The normalized cell lengths are given in the following corollary. These lengths are also shown in \figref{2BSBoundaries}.

\begin{corollary}
\label{cor:2BScellsize}
For $\Cmc = [-\Delta_1,+\Delta_2]$ with cell bias parameter $\beta$ and transmit power parameter $\tau$, the (normalized) cell lengths are functions of $\tilde{y}_{\pm} = \tilde{y}_{\pm}(\gamma(\sigma))$, for $\sigma = \sigma(\tau,\beta,\alpha)$:
\begin{equation}
\begin{array}{r|cc}
& |\Cmc_1| & |\Cmc_2| \\ \hline
i) & \tilde{y}_{+}-\tilde{y}_{-} & \Delta_1+\Delta_2 - (\tilde{y}_{+}-\tilde{y}_{-}) \nonumber \\
ii) & \tilde{y}_{+}+\Delta_1 & \Delta_2 - \tilde{y}_{+} \\
iii) & \Delta_1 & \Delta_2 \\
iv) & \tilde{y}_{-}+\Delta_1 & \Delta_2 - \tilde{y}_{-} \\
v) & \Delta_1+\Delta_2 - (\tilde{y}_{+}-\tilde{y}_{-}) & \tilde{y}_{+}-\tilde{y}_{-}
\end{array}
\end{equation}
where the five cases are: $i)$ $\sigma < -h(\Delta_1)$, $ii)$ $-h(\Delta_1) < \sigma < 0$, $iii)$ $\sigma = 0$, $iv)$ $0 < \sigma < +h(\Delta_2)$, and $v)$ $\sigma > +h(\Delta_2)$.
\end{corollary}
The cell lengths yield the probabilities that a randomly positioned user sits in a cell, with $p_i = |\Cmc_i|/|\Cmc|$, for $|\Cmc| = \Delta_1+\Delta_2$.

\subsection{Results}

\begin{figure*}
\centering
\includegraphics[width=5.7in]{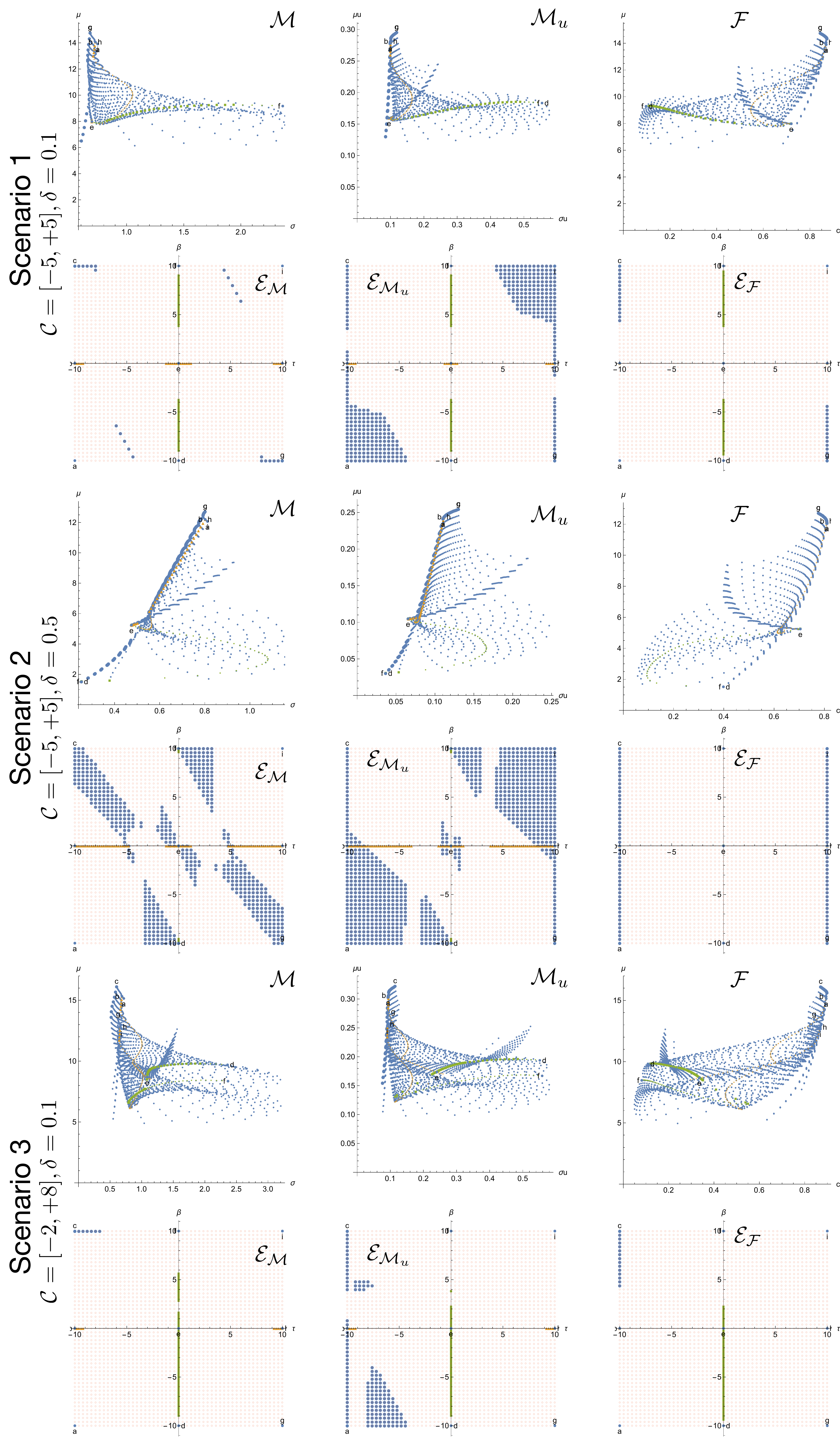}
\caption{(\secref{twobslinear}) SE tradeoffs for the two BS network.}
\label{fig:2BSresults}
\end{figure*}

\begin{figure}
\centering
\includegraphics[width=3.5in]{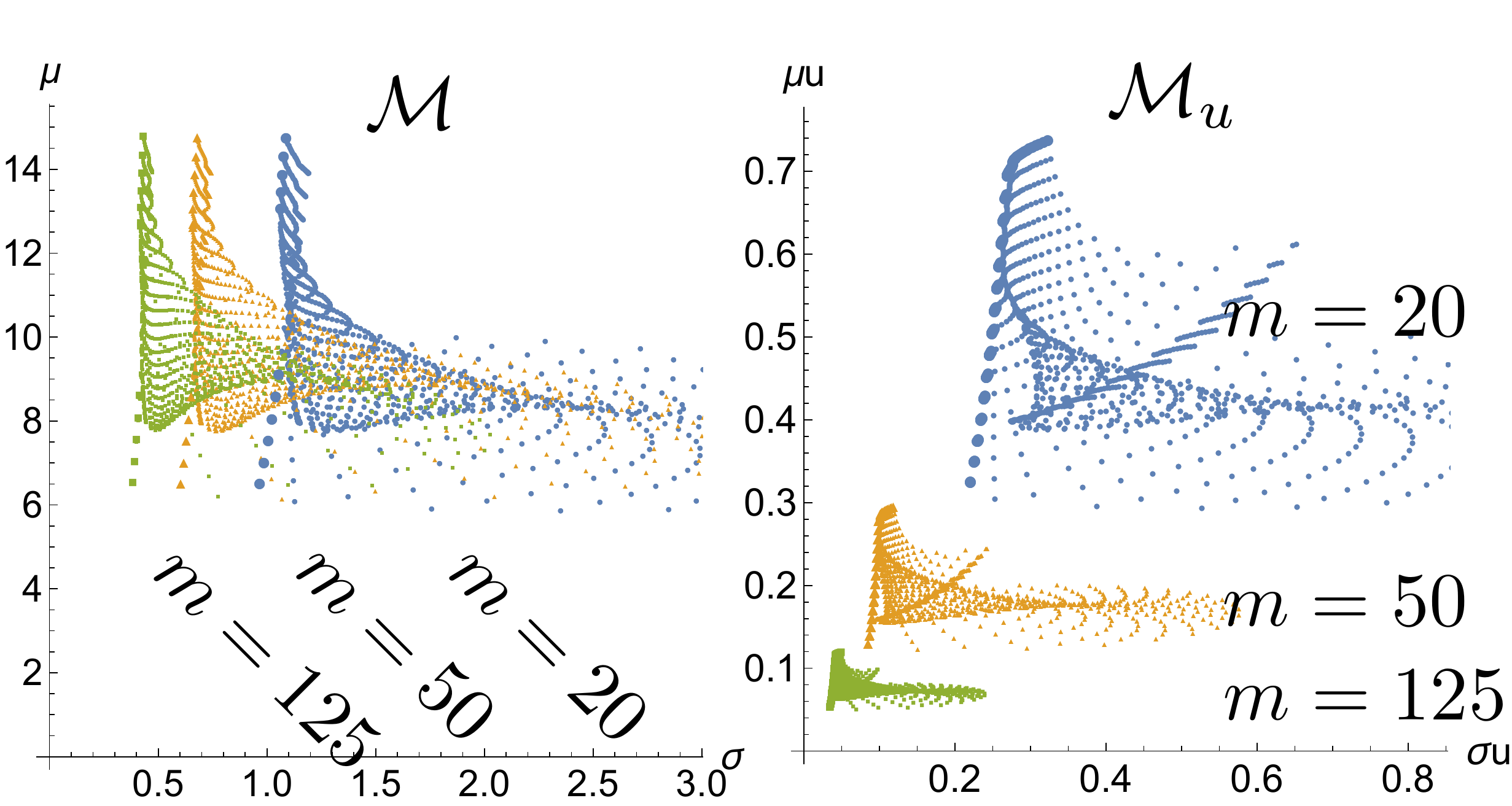}
\caption{(\secref{twobslinear}) $\Mmc$ (left) and $\Mmc_u$ (right) for $m \in \{20,50,125\}$.}
\label{fig:2BSresults2}
\end{figure}

SE tradeoffs for the two BS network are given in \figref{2BSresults} and \figref{2BSresults2}.  We first discuss \figref{2BSresults}.  Three scenarios are addressed: $i)$ $\Cmc = [-5,+5]$ and $\delta = 0.1$, $ii)$ $\Cmc = [-5,+5]$ and $\delta = 0.5$, and $iii)$ $\Cmc = [-2,+8]$ and $\delta = 0.1$.  Thus scenario $i)$ is a baseline of sorts, and scenario $ii)$ presents the impact of a larger deadzone radius $\delta$, while scenario $iii)$ presents the impact of an asymmetry in the network arena with respect to the two BS locations at $\pm 1$.  The top, middle, and bottom six plots in the $3 \times 6$ grid of plots in \figref{2BSresults} correspond to scenarios $i)$, $ii)$, and $iii)$, respectively.  Within each of the three scenarios we present the three tradeoffs $(\Mmc,\Mmc_u,\Fmc)$ (smaller symbols) and their Pareto efficient frontiers (larger symbols) on the top row (left to right), and the corresponding efficient controls $(\Emc_{\Mmc},\Emc_{\Mmc_u},\Emc_{\Fmc})$ on the bottom row (left to right).  Within each of the six plots we present results for $i)$ joint control of bias and power ($\Tmc \times \Bmc$, blue circles), $ii)$ control of power with bias fixed ($\Tmc \times \mathbf{1}$, orange triangles), and $iii)$ control of bias with power fixed ($\mathbf{1} \times \Bmc$, green squares).  In addition, we also show the nine extreme control points of $\Tmc \times \Bmc$ (labeled ``a'' through ``h'') and their corresponding performance in the tradeoff plot.  In all cases $\Tmc = [-10,+10]$ and $\Bmc = [-10,+10]$, each with a granularity of $51$ evenly spaced points, although we doubled this to $101$ points for $(\Tmc,\mathbf{1})$ and $(\mathbf{1},\Bmc)$.  We fixed $\alpha = 3$ and $m = 50$.  

The key observations from \figref{2BSresults} include the following.  First, the Pareto frontier for all three tradeoffs, $(\hat{\Mmc},\hat{\Mmc}_u,\hat{\Fmc})$ is {\em significantly} better under joint bias and power control $(\Tmc \times \Bmc)$ than under either power alone or bias alone.  Second, the set of efficient controls $(\Emc_{\Mmc},\Emc_{\Mmc_u},\Emc_{\Fmc})$ is {\em significantly} scenario dependent, meaning it is important to select the controls as a function of the network topology, here captured by the parameters $(\Delta_1,\Delta_2,\delta)$.  Third, given the choice between either power or bias (but not both), these results strongly suggest power control as the superior choice.  Fourth, under joint bias and power control, all three Pareto efficient frontiers  $(\hat{\Mmc},\hat{\Mmc}_u,\hat{\Fmc})$ are ``steep'', with a significant increase in $\mu,\mu_u,\bar{\mu}$ achievable by incurring a small cost in $\sigma,\sigma_u,\bar{c}$, respectively.  Fifth, the mapping from the control plane $(\tau,\beta)$ to the performance plane, e.g., $(\mu,\sigma)$, is non-trivial, as evidence by the, at least to us, difficult to predict mapping of the nine extreme points on the control plane.  Sixth, the set of efficient controls varies significantly across the three tradeoffs, so the control selection must be made with a particular tradeoff in mind.    

\figref{2BSresults2} shows $\Mmc$ (left) and $\Mmc_u$ (right) and their Pareto efficient frontiers for $m \in \{20,50,125\}$.  The figure illustrates the statements in \thmref{meanstddev} that $\mu_u^{(m)},\sigma^{(m)},\sigma_u^{(m)} \to 0$ and $\mu^{(m)} \to \bar{\mu}$ in $m$.  

%=========================================================================================
\section{Quincunx network}
\label{sec:quincunx}

\begin{figure}[!ht]
\centering
\includegraphics[width=3.5in]{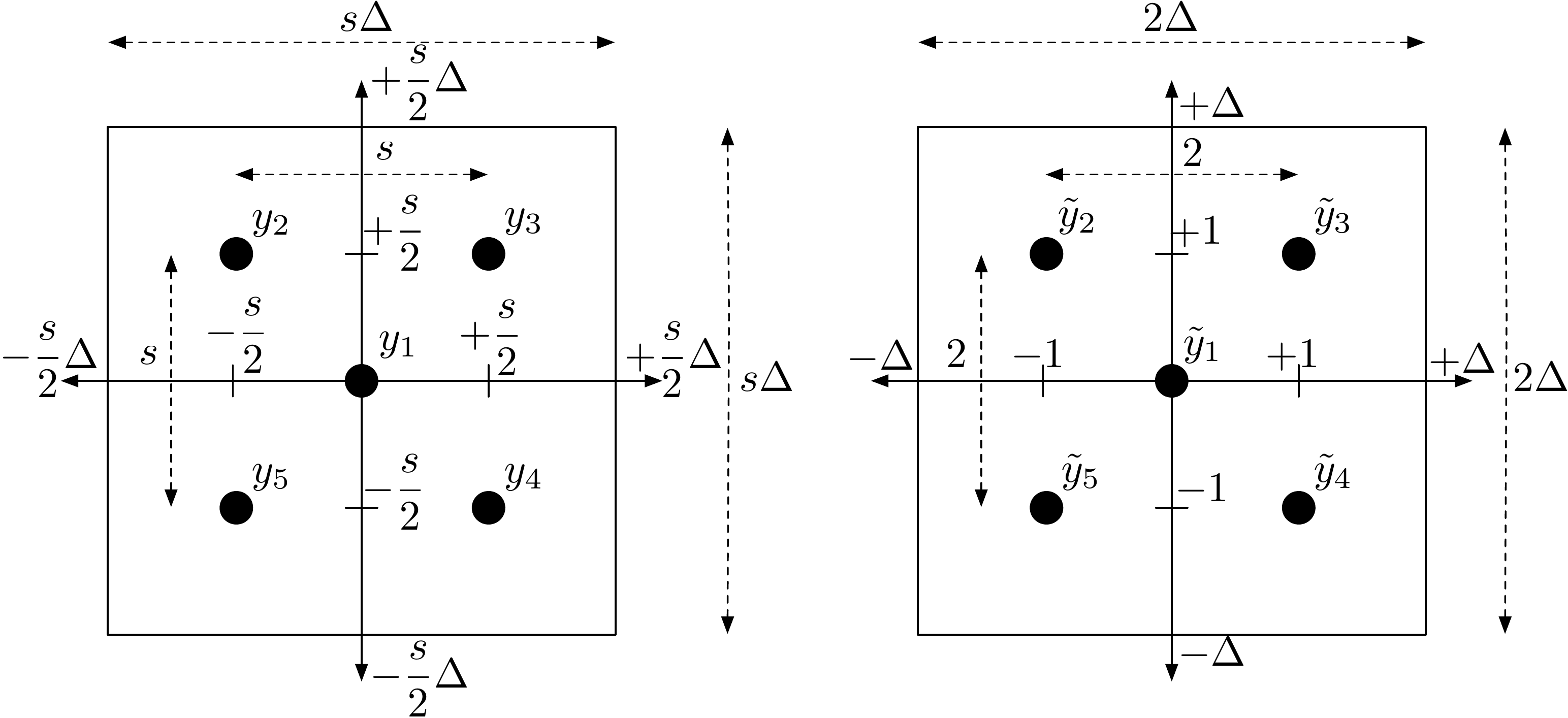}
\caption{(\secref{quincunx}) Five BS linear network with units of meters (left) and units of $s/2$ meters (right).}
\label{fig:5bstopology}
\end{figure}

Consider the two-dimensional cellular network with five BSs positioned as a quincunx \Cube{5}, with BS separation parameter $s>0$ and network arena parameter $\Delta > 1$, as shown in \figref{5bstopology}.  Using the same spatial scaling as in \secref{twobslinear}, a point $y = (y_{(1)},y_{(2)})$ maps to a point $\tilde{y} = (\tilde{y}_{(1)},\tilde{y}_{(2)})$ with each dimension scaled by $s/2$, as shown in \figref{5bstopology}.  In such units the network domain is $\Cmc = [-\Delta,+\Delta]^2$, with the central BS at $\tilde{y}_1 = (0,0)$ and the four corner BSs at $(\tilde{y}_2,\ldots,\tilde{y}_5) = \{(\pm1,\pm1)\}$.  As in \secref{twobslinear} we assume zero noise and add a ``dead-zone'' disk of radius $\delta$ around each BS.  Because $\|y-y'\|_2 = \frac{s}{2} \|\tilde{y}-\tilde{y}'\|_2$ and SIR depends upon ratios of distances, the analysis is independent of $s$. 

\begin{figure}[!ht]
\centering
\includegraphics[width=3.5in]{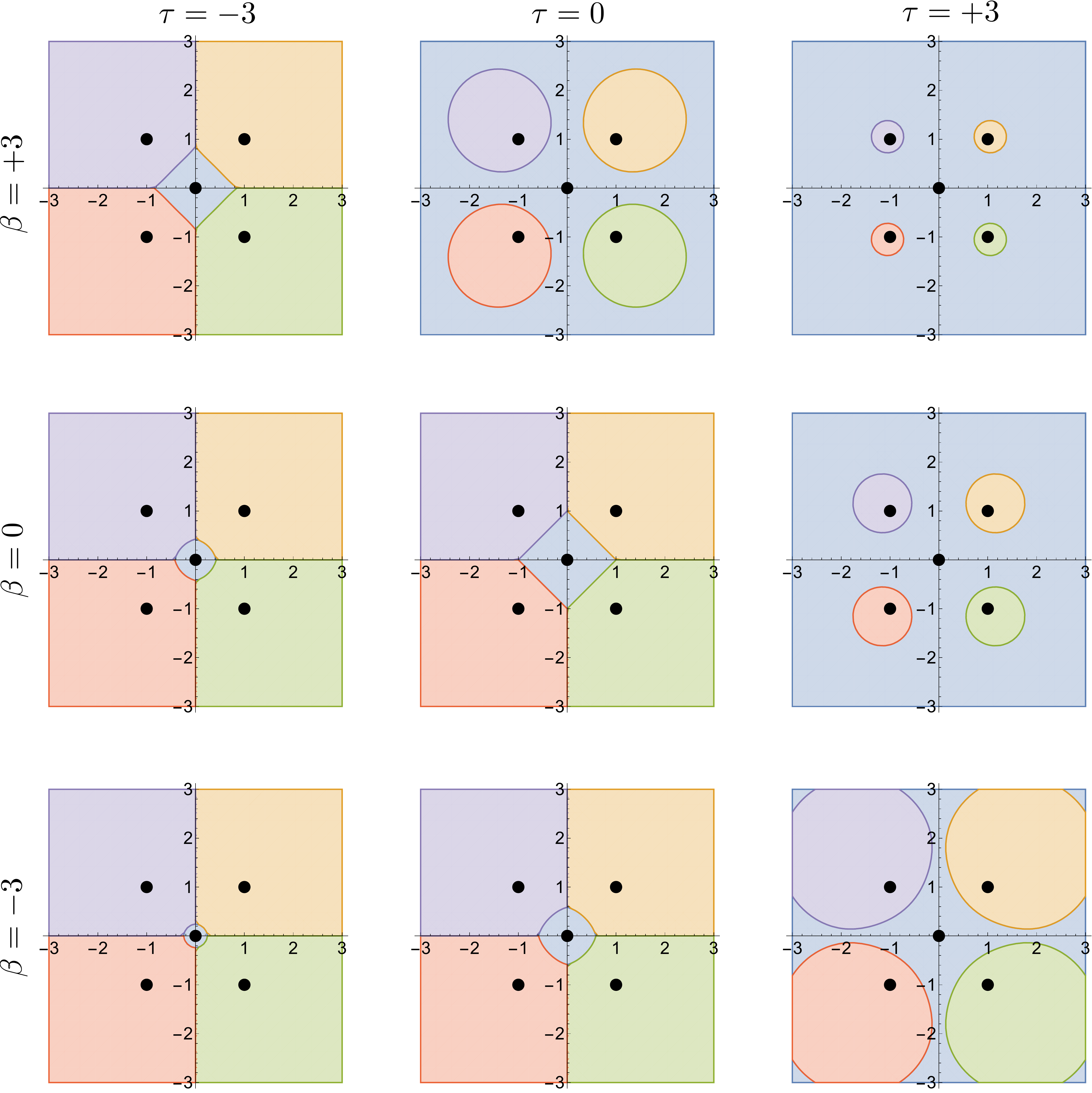}
\caption{(\secref{quincunx}) Cells $(\Cmc_1,\ldots,\Cmc_5)$ for $(\tau,\beta) \in \{-3,0,+3\}^2$.}
\label{fig:5bsregions}
\end{figure}

To simplify the design of the control space we restrict to the case where the four corner BSs use a common transmit power and a common bias parameter, i.e., $t_i = t_c$ and $b_i = b_c$ for parameters $(t_c,b_c)$ and $i \in \{2,3,4,5\}$.  In the no noise regime, the above restriction means the effects of both bias and transmit power are determined by $t_1/t_c$ and $b_1/b_c$.  This allows reparameterization of the control space in terms of $(\tau,\beta)$, with 
\begin{equation}
\tau = \log \left( \frac{t_1}{t_c} \right), ~ \beta = \log \left( \frac{b_1}{b_c} \right).
\end{equation}
The cells $(\Cmc_1,\ldots,\Cmc_5)$ for the nine combinations $(\tau,\beta) \in \{-3,0,+3\}^2$ with $\Delta = 3$ and $\alpha = 3$ are shown in \figref{5bsregions}.

\begin{figure*}
\centering
\includegraphics[width=6.5in]{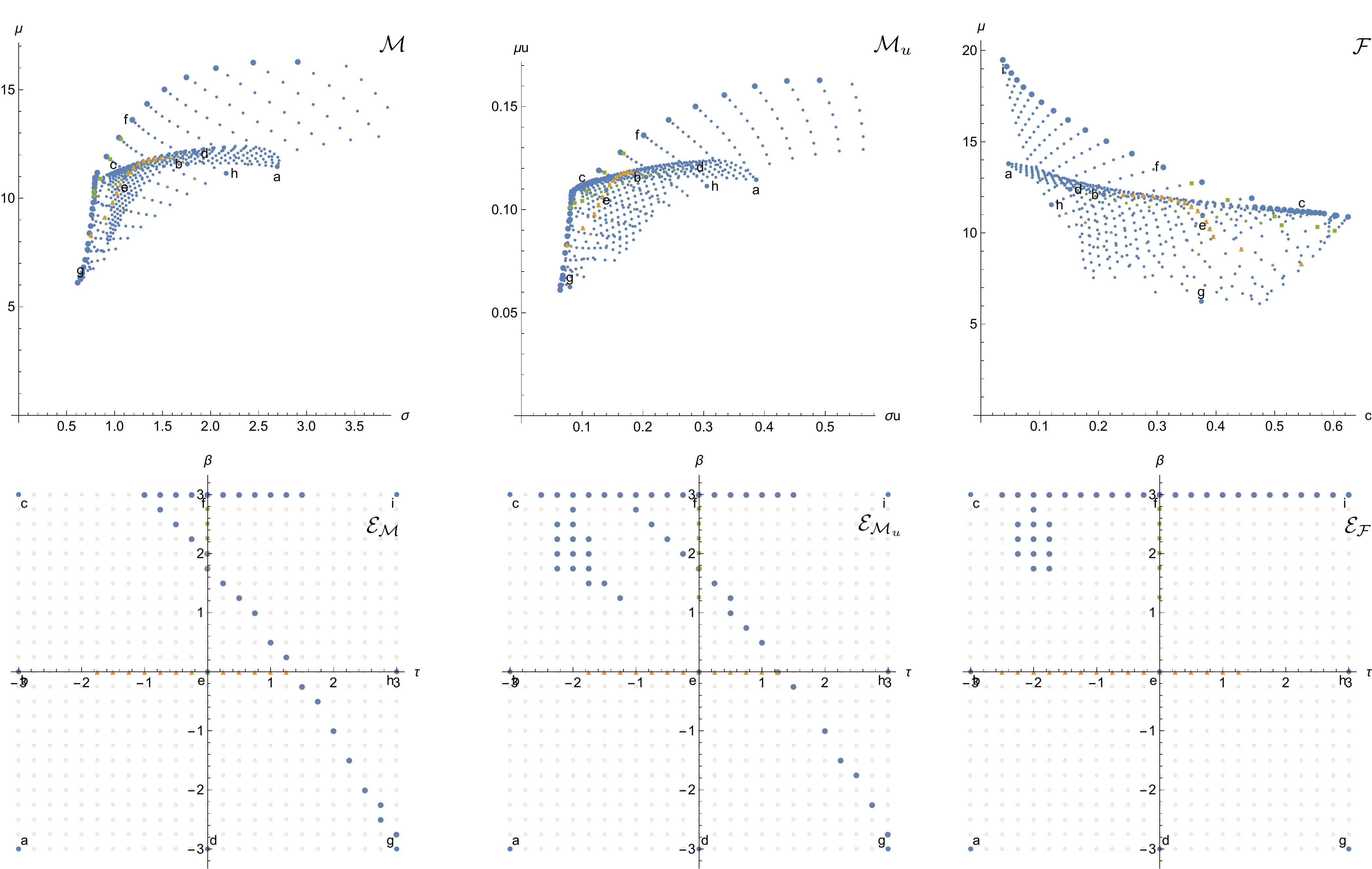}
\caption{(\secref{quincunx}) SE tradeoffs for the five BS network.}
\label{fig:5bsresults}
\end{figure*}

Numerical results are shown in \figref{5bsresults}.  As in \figref{2BSresults}, the plots show tradeoffs $(\Mmc,\Mmc_u,\Fmc)$ (top row, left to right), and the corresponding efficient controls $(\Emc_{\Mmc},\Emc_{\Mmc_u},\Emc_{\Fmc})$ (bottom row, left to right), for $i)$ joint bias and power $\Tmc \times \Bmc$ (blue circles), $ii)$ power $\Tmc \times \mathbf{1}$ (orange triangles), and $iii)$ bias $\mathbf{1} \times \Bmc$ (green squares).  Efficient frontiers $(\hat{\Mmc},\hat{\Mmc}_u,\hat{\Fmc})$ are larger symbols, dominated points are smaller symbols.  The nine extreme points of the control space are also shown.  The parameters are: $\Tmc = [-3,+3]$, and $\Bmc = [-3,+3]$, each quantized at $25$ evenly spaced points, and $\alpha = 3$, $\delta=0$, $m=100$.  

Two observations, aside from those already made regarding \figref{2BSresults}, warrant mention.  First, bias alone is, for this topology, a better univariate control than is power alone, to achieve the Pareto frontiers for $\Mmc,\Mmc_u$; this is in contrast to what is observed in \figref{2BSresults}, where power alone was superior to bias alone in several cases.  Second, the efficient controls $\Emc_{\Mmc},\Emc_{\Mmc_u}$ demonstrate part of the Pareto frontier of $\Mmc,\Mmc_u$ is achieved by selecting bias to be maximum and sweeping power, and another part of the frontier is achieved by selecting power and bias to be inverse to one another.  In the latter case we have, say, the center cell using a low bias and high power relative to the corner cells, which creates a very high SE for the users in the small center cell, and comparatively low SE for the users in the larger corner cells.  Sweeping $(\tau,\beta)$ to have a linear constant sum roughly traces out the efficient frontier $\hat{\Mmc}$.  

%=========================================================================================
\section{Conclusions and future work}
\label{sec:conclusions}

The primary contributions of this work are $i)$ explicit expressions for performance metrics $(\mu,\sigma,\mu_u,\sigma_u,\bar{\mu},\bar{c})$ related to the collection of per-user SEs in terms of the statistics $(\pbf,\psi,\psi^{(2)})$, which in turn depend upon the bias and power controls $(\tbf,\bbf)$, and $ii)$ demonstration that joint bias and power control achieves (in some but not all cases) a significant performance benefit in terms of SE tradeoffs $(\Mmc,\Mmc_u,\Fmc)$ over either control alone.  A natural, perhaps ambitious, next step is to analytically characterize efficient frontiers and controls.

%=========================================================================================
\appendix

%=========================================================================================
\subsection{Proof of \thmref{meanstddev} (SE mean and standard deviation)}
\label{prf:meanstddev}

Before proving \thmref{meanstddev} we first establish \lemref{covmimj} and \lemref{exp1m}.  

\begin{lemma}
\label{lem:covmimj}
Let $\Mbf = (M_1,\ldots,M_n) \sim \mathrm{mult}(m,\pbf)$ be a multinomial RV as in \eqref{multdisbn}, and let $\mathbf{1}_{M_i > 0}$ be a RV indicating whether or not cell $i$ is occupied.  Then 
\begin{eqnarray}
\!\!\!\!\!\!\!\!\! \Pbb(M_i > 0, M_j > 0) \!\!\!\!\!&=&\!\!\!\!\! 1-(\bar{p}_i^m + \bar{p}_j^m) + (1-(p_i + p_j))^m \label{eq:pmimj1} \\
\mathrm{Cov}(\mathbf{1}_{M_i > 0},\mathbf{1}_{M_j > 0}) \!\!\!\!\!&=&\!\!\!\!\! (1-(p_i+p_j))^m - (\bar{p}_i \bar{p}_j)^m \label{eq:covindmimj}
\end{eqnarray}
\end{lemma}

\begin{IEEEproof}
We first establish \eqref{pmimj1}.  Rewrite the left side as  
\begin{equation}
\label{eq:pmimj2}
\Pbb(M_i > 0, M_j > 0) = \Pbb(M_i > 0| M_j > 0)\Pbb(M_j > 0)
\end{equation}
and introduce shorthand notation $P_{i,j} = P_{i|j} P_j$ for the above equation. Given $M_j = v$, the RV $M_i|_{M_j = v}$ has a binomial distribution $\mathrm{bin}(m-v,p_i/(1-p_j))$.  This allows
\begin{eqnarray}
P_{i|j}
&=& \sum_{v=1}^m \Pbb(M_i > 0 | M_j > 0, M_j = v) \Pbb(M_j = v | M_j > 0) \nonumber \\
&=& \frac{1}{P_j} \sum_{v=1}^m \Pbb(M_i > 0 | M_j = v) \Pbb(M_j = v) \nonumber \\
&=& \frac{1}{P_j} \sum_{v=1}^m \left(1 - \left(1 - \frac{p_i}{1-p_j}\right)^{m-v} \right)\Pbb(M_j = v) \nonumber \\
&=& 1 - \frac{1}{P_j} \sum_{v=1}^m \left(1 - \frac{p_i}{1-p_j}\right)^{m-v} \Pbb(M_j = v) \nonumber \\
&=& 1 - \frac{1}{P_j} \sum_{v=1}^m \binom{m}{v} (1-(p_i+p_j))^{m-v} p_j^v  \nonumber \\
&=& 1 - \frac{1}{P_j} \left( (1-p_i)^m - (1-(p_i+p_j))^m \right) 
\end{eqnarray}
where the last step is the binomial theorem.  Substituting into \eqref{pmimj2} gives \eqref{pmimj1}.  We obtain \eqref{covindmimj} by substituting \eqref{pmimj1} into $\mathrm{Cov}(\mathbf{1}_{M_i > 0},\mathbf{1}_{M_j > 0}) =$  
\begin{equation}
\Ebb[\mathbf{1}_{M_i > 0}\mathbf{1}_{M_j > 0}] - \Ebb[\mathbf{1}_{M_i > 0}] \Ebb[\mathbf{1}_{M_j > 0}] = P_{ij} - P_i P_j.
\end{equation}
\end{IEEEproof}

\begin{lemma}
\label{lem:exp1m}
Let $M^{(m)} \sim \mathrm{bin}(m,p)$ be a binomial RV, and define $I^{(m)} = \frac{1}{M^{(m)}} \mathbf{1}_{M^{(m)}>0}$.  Then $\lim_{m \to \infty} \Ebb[I^{(m)}] = 0$.  
\end{lemma}

\begin{IEEEproof}
It is easy to see that the function 
\begin{equation}
h_r^{(m)}(k) = \left\{ \begin{array}{ll}
1, \; & 1 \leq k < r \\
1/r, \; & r \leq k \leq m 
\end{array} \right.
\end{equation}
with parameter $r \in [m]$ satisfies $1/k \leq h_r^{(m)}(k)$ for all $k \in [m]$.  This allows us to split the sum in the expectation at $r$:
\begin{eqnarray}
\Ebb[I^{(m)}] 
&=& \sum_{k=1}^m \frac{1}{k} \Pbb(M =k) \nonumber \\
&\leq& \sum_{k=1}^m h_r^{(m)}(k) \Pbb(M =k) \nonumber \\
&=& \Pbb(M < r) + \frac{1}{r} \Pbb(M \geq r) \nonumber \\
&=& \frac{1}{r} + \left(1 - \frac{1}{r} \right) \Pbb(M < r) 
\end{eqnarray}
A commonly used Chernoff bound on the binomial tail probability is, for $M \sim \mathrm{bin}(m,p)$,
\begin{equation}
\Pbb(M \leq r) \leq \left( \frac{m(1-p)}{m-r} \right)^{m-r} \left( \frac{mp}{r} \right)^r, 
\end{equation}
for any $r \leq m p$.  We write $r = \delta m p$, for $\delta \in (0,1)$, to yield
\begin{equation}
\Pbb(M \leq \delta m p) \leq \left( \left( \frac{1-p}{1-\delta p} \right)^{1-\delta p} \left( \frac{1}{\delta} \right)^{\delta p} \right)^m.
\end{equation}
Substitution gives the following upper bound on $\Ebb[I^{(m)}]$, denoted as $\tilde{g}(m;\delta)$, defined as
\begin{equation}
\tilde{g}(m;\delta) \equiv \frac{1}{\delta m p} + \left(1 - \frac{1}{\delta m p} \right) \left( \left( \frac{1-p}{1-\delta p} \right)^{1-\delta p} \left( \frac{1}{\delta} \right)^{\delta p} \right)^m.
\end{equation}
The optimal $\delta$ as a function of $m$ and $p$ to minimize $\tilde{g}$ is difficult to obtain, but is not required, as $\delta = 1/2$ suffices:
\begin{equation}
\tilde{g}(m;1/2) = \frac{2}{m p} + \left(1 - \frac{2}{m p} \right) \left( \left( \frac{1-p}{1-p/2} \right)^{1-p/2} 2^{p/2} \right)^m 
\end{equation}
One can show that 
\begin{equation}
p \in (0,1) \Rightarrow \left( \frac{1-p}{1-p/2} \right)^{1-p/2} 2^{p/2} \in (0,1),
\end{equation}
which ensures $\lim_{m \to \infty} \tilde{g}(m;1/2) = 0$.  The required limit follows as $\Ebb[I^{(m)}] \leq \tilde{g}(m;\delta)$.  
\end{IEEEproof}

We now proceed to the proof of \thmref{meanstddev}.
\begin{IEEEproof}
The proof is divided into four steps: $i)$ $\mu^{(m)}$ and $\bar{\mu}$, $ii)$ $\sigma^{(m)}$ and $\bar{\sigma}$, $iii)$ $\mu_u^{(m)}$ and $\bar{\mu}_u$, and $iv)$ $\sigma^{(m)}_u$ and $\bar{\sigma}_u$.  At the risk of slight ambiguity, we suppress the superscript $(m)$ for finite $m$ to make the notation less burdensome.  In what follows we write $\Vrm(\cdot),\Crm(\cdot,\cdot)$ as shorthand for $\mathrm{Var}(\cdot),\mathrm{Cov}(\cdot,\cdot)$.  

\underline{Step $i)$:} $\mu^{(m)}$ and $\bar{\mu}$.  We will use the property of conditional expectation $\Ebb[X] = \Ebb_{\Mbf}[\Ebb[X|\Mbf]]$, where $X = X^{(m)}$ is the random total SE and $\Mbf = \Mbf^{(m)}$ is the random occupancy count vector.  Then:
\begin{eqnarray}
\Ebb[X|\Mbf] 
&=& \Ebb \left[ \left. \sum_{u=1}^m X_u \right| \Mbf \right] \nonumber \\
&=& \Ebb \left[ \left. \sum_{i \in \Amc(\Mbf)} \sum_{u \in \Umc_i} X_u \right| \Mbf \right] \nonumber \\
&=& \sum_{i \in \Amc(\Mbf)} \sum_{u \in \Umc_i} \Ebb[X_u|M_i,i(u)=i]  
\end{eqnarray}
where $i(u) \in [n]$ is the random BS assignnment for user $u$'s random location $Z_u$: $Z_u \in \Cmc_{i(u)}$.  Then $\Ebb[X_u|M_i,i(u)=i] = \frac{\psi_i}{M_i}$ and substitution gives 
\begin{equation}
\Ebb[X|\Mbf] = \sum_{i \in \Amc(\Mbf)} \sum_{u \in \Umc_i} \frac{\psi_i}{M_i} = \sum_{i \in \Amc(\Mbf)} \psi_i.
\end{equation}
Finally, 
\begin{equation}
\mu^{(m)} = \Ebb_{\Mbf}\left[ \sum_{i \in \Amc(\Mbf)} \psi_i \right] = \sum_{i=1}^n \psi_i \Pbb(M_i > 0) = \sum_{i=1}^n (1-\bar{p}_i^m) \psi_i.
\end{equation}
The limit $\bar{\mu} = \lim_{m \to \infty} \mu^{(m)}$ follows directly.

\underline{Step $ii)$:} $\sigma^{(m)}$ and $\bar{\sigma}$.  We use the law of total variance:
\begin{equation}
\label{eq:thm1prfstep2lotv}
\Vrm(X) = \Ebb_{\Mbf}\left[ \Vrm(X|\Mbf) \right] + \Vrm_{\Mbf}(\Ebb[X|\Mbf]).
\end{equation}
We will address the two terms in the above sum separately.  

First term in \eqref{thm1prfstep2lotv}: define $\Pmc(\mbf) = \{ 1 \leq i < j \leq n : m_i > 0 \mbox{ and } m_j > 0\}$ as the set of distinct unordered pairs of occupied cells.  For the variance in the first term we use the equation for the variance of a sum of RVs: $\Vrm(X|\Mbf)$
\begin{eqnarray}
&=& \sum_{u=1}^m \Vrm(X_u|\Mbf) + 2 \sum_{1 \leq u < v \leq m} \Crm(X_u,X_v|\Mbf) \\
&=& \sum_{u=1}^m \Vrm(X_u|M_{i(u)}) + 2 \!\!\!\!\! \sum_{1 \leq u < v \leq m} \!\!\!\!\!\Crm(X_u,X_v|M_{i(u)},M_{i(v)}) \nonumber \\
&=& \sum_{i \in \Amc(\Mbf)} \sum_{u \in \Umc_i} \Vrm(X_u|M_i,i(u)=i) \nonumber \\
& + &  2 \sum_{i \in \Amc(\Mbf)} \sum_{\{u,v\} \in \Umc_i^2} \Crm(X_u,X_v|M_i,i(u)=i,i(v)=i) \nonumber \\
& + & \!\!\!\!\! 2 \!\!\!\!\! \!\!\!\!\! \sum_{\{i,j\} \in \Pmc(\Mbf)} \sum_{u \in \Umc_i} \sum_{v \in \Umc_j} \Crm(X_u,X_v|M_i,M_j,i(u)=i,i(v)=j) \nonumber  
\end{eqnarray}
We now proceed to study each of the three types of terms in the above sum.  First: $\Vrm(X_u|M_i,i(u)=i)$
\begin{equation}
= \Ebb[X_u^2|M_i,i(u)=i] - \Ebb[X_u|M_i,i(u)=i]^2 
= \frac{\psi^{(2)}_i - \psi_i^2}{M_i^2}.
\end{equation}
Second: $\Crm(X_u,X_v|M_i,i(u)=i,i(v)=i) = 0$ since $X_u,X_v$ with $i(u) = i(v) = i$ are conditionally independent given $M_i$. Third: similarly, $\Crm(X_u,X_v|M_i,M_j,i(u)=i,i(v)=j) = 0$ since $X_u,X_v$ with $i(u)=i,i(v)=j$ are conditionally independent given $M_i,M_j$.  Thus: $\Vrm(X|\Mbf)$
\begin{equation}
= \sum_{i \in \Amc(\Mbf)} \sum_{u \in \Umc_i}\Vrm(X_u|M_i,i(u)=i)
= \sum_{i \in \Amc(\Mbf)} \frac{\psi^{(2)}_i - \psi_i^2}{M_i}
\end{equation}
We now complete the derivation of the first term in \eqref{thm1prfstep2lotv}: 
\begin{eqnarray}
\Ebb_{\Mbf}[\mathrm{Var}(X|\Mbf)] 
&=& \Ebb_{\Mbf} \left[ \sum_{i \in \Amc(\Mbf)} \frac{\psi^{(2)}_i - \psi_i^2}{M_i} \right] \label{eq:thm1prfstep2firstterm} \\ 
&=& \sum_{i=1}^n (\psi^{(2)}_i - \psi_i^2) \Ebb\left[\frac{1}{M_i} \mathbf{1}_{M_i > 0} \right] \nonumber 
\end{eqnarray}
Second term in \eqref{thm1prfstep2lotv}: we find $\Vrm_{\Mbf}(\Ebb[X|\Mbf])$ using the equation for the variance of a weighted sum of RVs: $\Vrm_{\Mbf}(\Ebb[X|\Mbf])$
\begin{eqnarray}
&=& \Vrm_{\Mbf} \left( \sum_{i=1}^n \psi_i \mathbf{1}_{M_i > 0} \right) \label{eq:thm1prfstep2vws} \\
&=& \sum_{i=1}^n \psi_i^2 \Vrm(\mathbf{1}_{M_i > 0}) + 2 \!\!\!\!\! \sum_{1 \leq i < j \leq n} \psi_i \psi_j \Crm(\mathbf{1}_{M_i > 0},\mathbf{1}_{M_j > 0}) \nonumber
\end{eqnarray}
The variances in the first sum in \eqref{thm1prfstep2vws} are: $\Vrm(\mathbf{1}_{M_i > 0})$
\begin{equation}
 = \Pbb(M_i>0)(1-\Pbb(M_i > 0)) = (1-\bar{p}_i^m)\bar{p}_i^m.
\end{equation}
The covariances in the second sum in \eqref{thm1prfstep2vws} are given by \eqref{covindmimj} in \lemref{covmimj}.  Combining: $\Vrm_{\Mbf}(\Ebb[X|\Mbf])$
\begin{eqnarray}
&=& \sum_{i=1}^n \psi_i^2 (1-\bar{p}_i^m)\bar{p}_i^m \label{eq:thm1prfstep2secondterm} \\
&+& 2 \sum_{1 \leq i < j \leq n} \psi_i \psi_j \left( (1-(p_i+p_j))^m - (\bar{p}_i \bar{p}_j)^m \right) \nonumber 
\end{eqnarray}
Substitution of \eqref{thm1prfstep2firstterm} and \eqref{thm1prfstep2secondterm} into \eqref{thm1prfstep2lotv} gives \eqref{varm}.  It remains to show $\bar{\sigma}^2 = \lim_{m \to \infty} (\sigma^{(m)})^2 = 0$.  The second and third terms in \eqref{varm} clearly converge to $0$, so it suffices to show $\Ebb[\mathbf{1}_{M_i > 0}/M_i] \to 0$ as $m \to \infty$.  This is shown in \lemref{exp1m}. 

\underline{Step $iii)$:} $\mu_u^{(m)}$ and $\bar{\mu}_u$.  Denote the randomly selected user index as the random variable $U \in [m]$.  Conditioned on the occupancy vector $\Mbf$, we have $\Pbb(U \in \Umc_i|\Mbf) = M_i/m$.  Thus:
\begin{eqnarray}
\Ebb[X_U|\Mbf]
&=& \sum_{i \in \Amc(\Mbf)} \Ebb[X_U|\Mbf, U \in \Umc_i]\Pbb(U \in \Umc_i|\Mbf) \nonumber \\
&=& \sum_{i \in \Amc(\Mbf)} \frac{\psi_i}{M_i} \frac{M_i}{m}  = \frac{1}{m} \sum_{i \in \Amc(\Mbf)} \psi_i
\end{eqnarray}
and
\begin{eqnarray}
\Ebb[X_U] &=& \Ebb_{\Mbf}[ \Ebb[X_U|\Mbf]] \nonumber \\
&=& \frac{1}{m} \Ebb_{\Mbf} \left[ \sum_{i=1}^n \psi_i \mathbf{1}_{M_i > 0} \right] \nonumber \\
&=& \frac{1}{m} \sum_{i=1}^n \psi_i \Pbb(M_i > 0) \nonumber \\
&=& \frac{1}{m} \sum_{i=1}^n (1-\bar{p}_i^m) \psi_i = \frac{\mu^{(m)}}{m} 
\end{eqnarray}
The fact that $\mu^{(m)}_u \to 0$ follows from $\mu^{(m)} \to \bar{\mu}$.

\underline{Step $iv)$:} $\sigma^{(m)}_u$ and $\bar{\sigma}_u$.  We follow a similar  strategy used in the proof of \underline{Step $ii)$}, above.  Denote the randomly selected user index as the RV $U \in [m]$; the law of total variance gives
\begin{equation}
\label{eq:thm1prfstep4lotv}
\Vrm(X_U) = \Ebb_{\Mbf}\left[ \Vrm(X_U|\Mbf) \right] + \Vrm_{\Mbf}(\Ebb[X_U|\Mbf]).
\end{equation}
We will address the two terms in the above sum separately.

First term in \eqref{thm1prfstep4lotv}: to find $\Ebb_{\Mbf}\left[ \Vrm(X_U|\Mbf) \right]$ we first find $\Vrm(X_U|\Mbf)$.  To find $\Vrm(X_U|\Mbf)$ we use a conditional form of the law of total variance, conditioned on $\Mbf$:
\begin{equation}
\Vrm(X_U|\Mbf) = \Ebb_U[ \Vrm(X_U|U,\Mbf) |\Mbf] + \Vrm_U( \Ebb[X_U|U,\Mbf] | \Mbf ).
\label{eq:thm1prfstep4clotv}
\end{equation}
We address the two terms in \eqref{thm1prfstep4clotv} separately.  
First term in \eqref{thm1prfstep4clotv}: to find $\Ebb_U[ \Vrm(X_U|U,\Mbf) |\Mbf]$ we first find $\Vrm(X_U|U,\Mbf)$:
\begin{equation}
= \Ebb[X_U^2|U,M_U] - \Ebb[X_U|U,M_U]^2 
= \frac{\psi^{(2)}_{i(U)} - \psi_{i(U)}^2}{M_{i(U)}^2}
\end{equation}
Thus: $\Ebb_U[ \Vrm(X_U|U,\Mbf) |\Mbf]$
\begin{eqnarray}
&=& \Ebb_U \left[ \left. \frac{\psi^{(2)}_{i(U)} - \psi_{i(U)}^2}{M_{i(U)}^2} \right| \Mbf \right] \nonumber \\
&=& \sum_{i \in \Amc(\Mbf)} \Ebb_U \left[ \left. \frac{\psi^{(2)}_{i(U)} - \psi_{i(U)}^2}{M_{i(U)}^2} \right| U \in \Umc_i,\Mbf \right] \Pbb(U \in \Umc_i | \Mbf) \nonumber \\
&=& \frac{1}{m} \sum_{i \in \Amc(\Mbf)} \frac{\psi^{(2)}_i - \psi_i^2}{M_i} \label{eq:thm1prfstep4term11}
\end{eqnarray}
Second term in \eqref{thm1prfstep4clotv}: to find $\Vrm_U( \Ebb[X_U|U,\Mbf] | \Mbf )$ we first find $\Ebb[X_U|U,\Mbf] = \psi_{i(U)}/M_{i(U)}$, and thus $\Vrm_U( \Ebb[X_U|U,\Mbf] | \Mbf)$
\begin{eqnarray} 
&=& \Vrm_U \left( \left. \frac{\psi_{i(U)}}{M_{i(U)}} \right| \Mbf \right) \nonumber \\
&=& \Ebb_U \left[ \left. \left( \frac{\psi_{i(U)}}{M_{i(U)}} \right)^2 \right| \Mbf \right] - \Ebb_U \left[ \left. \frac{\psi_{i(U)}}{M_{i(U)}} \right| \Mbf \right]^2 \nonumber \\
&=& \sum_{i \in \Amc(\Mbf)} \left( \frac{\psi_i}{M_i} \right)^2 \frac{M_i}{m} - \left( \sum_{i \in \Amc(\Mbf)} \frac{\psi_i}{M_i} \frac{M_i}{m} \right)^2 \nonumber \\
&=& \frac{1}{m} \sum_{i \in \Amc(\Mbf)} \frac{\psi_i^2}{M_i} - \frac{1}{m^2} \left( \sum_{i \in \Amc(\Mbf)} \psi_i \right)^2 \label{eq:thm1prfstep4term12}
\end{eqnarray}
To complete the first term in \eqref{thm1prfstep4lotv} we take the expectation of the sum of \eqref{thm1prfstep4term11} and \eqref{thm1prfstep4term12}: $\Ebb_{\Mbf}\left[ \Vrm(X_U|\Mbf) \right]$
\begin{eqnarray}
&=& \Ebb_{\Mbf}\left[ \Ebb_U[ \Vrm(X_U|U,\Mbf) |\Mbf] + \Vrm_U( \Ebb[X_U|U,\Mbf] | \Mbf ) \right] \nonumber \\
&=& \frac{1}{m} \Ebb\left[ \sum_{i=1}^n \frac{\psi^{(2)}_i - \psi_i^2}{M_i} \mathbf{1}_{M_i > 0} \right] + \frac{1}{m} \Ebb \left[ \sum_{i=1}^n \frac{\psi_i^2}{M_i} \mathbf{1}_{M_i > 0} \right] \nonumber \\
&-& \frac{1}{m^2} \Ebb \left[ \left( \sum_{i=1}^n \psi_i \mathbf{1}_{M_i > 0} \right)^2 \right] \nonumber \\
&=& \frac{1}{m} \Ebb\left[ \sum_{i=1}^n \psi^{(2)}_i \frac{1}{M_i} \mathbf{1}_{M_i > 0} \right] - \frac{1}{m^2} \Ebb\left[ \sum_{i=1}^n \psi_i^2 \mathbf{1}_{M_i > 0} \right] \nonumber \\ 
&-& \frac{2}{m^2} \Ebb\left[ \sum_{1 \leq i < j \leq n} \psi_i \psi_j \mathbf{1}_{M_i > 0} \mathbf{1}_{M_j > 0} \right] \nonumber \\
&=& \frac{1}{m} \sum_{i=1}^n \psi^{(2)}_i \Ebb\left[ \frac{1}{M_i} \mathbf{1}_{M_i > 0} \right] - \frac{1}{m^2} \sum_{i=1}^n \psi_i^2 (1-\bar{p}_i^m) \nonumber \\
&-& \frac{2}{m^2} \sum_{1 \leq i < j \leq n} \psi_i \psi_j ((1-(p_i+p_j))^m - (\bar{p}_i \bar{p}_j)^m) \label{eq:thm1prfstep41sttermfinal}
\end{eqnarray}
The expression for $\Pbb(M_i > 0, M_j > 0)$ is from \eqref{pmimj1} in \lemref{covmimj}.

Second term in \eqref{thm1prfstep4lotv}: to find $\Vrm_{\Mbf}(\Ebb[X_U|\Mbf])$ we first find
\begin{eqnarray}
\Ebb[X_U|\Mbf] 
&=& \Ebb_U [ \Ebb[ X_U | U,\Mbf ] | \Mbf] \nonumber \\ 
&=& \Ebb_U \left[ \left. \frac{\psi_{i(U)}}{M_{i(U)}} \right| \Mbf \right] \nonumber \\
&=& \sum_{i \in \Amc(\Mbf)} \frac{\psi_i}{M_i} \frac{M_i}{m} = \frac{1}{m} \sum_{i \in \Amc(\Mbf)} \psi_i 
\end{eqnarray}
and thus $\Vrm_{\Mbf}(\Ebb[X_U|\Mbf])$
\begin{eqnarray}
&=& \frac{1}{m^2} \sum_{i=1}^n \psi_i^2 \Vrm(\mathbf{1}_{M_i > 0}) \nonumber \\
&+& \frac{2}{m^2}  \sum_{1 \leq i < j \leq n} \psi_i \psi_j \Crm(\mathbf{1}_{M_i > 0},\mathbf{1}_{M_i > 0}) \nonumber \\
&=& \frac{1}{m^2} \sum_{i=1}^n \psi_i^2 (1-\bar{p}_i^m)\bar{p}_i^m \label{eq:thm1prfstep42ndtermfinal} \\
&+& \frac{2}{m^2} \sum_{1 \leq i < j \leq n} \psi_i \psi_j \left( (1-(p_i+p_j))^m - (\bar{p}_i \bar{p}_j)^m \right) \nonumber 
\end{eqnarray}
We obtain \eqref{sigum} by substituting \eqref{thm1prfstep41sttermfinal} and \eqref{thm1prfstep42ndtermfinal} into \eqref{thm1prfstep4lotv} and simplifying.  To see that $\sigma^{(m)}_u \to \bar{\sigma}_u = 0$, observe that each of the three terms in \eqref{sigum} each converge to $0$ as $m \to \infty$, where the convergence of the first term follows from \lemref{exp1m}.
\end{IEEEproof}

%=========================================================================================
\subsection{Proof of \thmref{fairness} (SE asymptotic fairness)}
\label{prf:fairness}

We prove \lemref{expmMconv}, \lemref{varmMto0}, and \lemref{covmMto0} before \thmref{fairness}.

\begin{lemma}
\label{lem:expmMconv}
Let $M^{(m)} \sim \mathrm{bin}(m,p)$ be a binomial RV, and define $I^{(m)} = \frac{1}{M^{(m)}} \mathbf{1}_{M^{(m)}>0}$.  Then $f^{(m)} = \Ebb \left[m I^{(m)} \right] \to \frac{1}{p}$.  
\end{lemma} 

\begin{IEEEproof}
We establish the recurrence
\begin{equation}
\label{eq:exp1mrec}
f^{(m)} = 1 - \bar{p}^m + \frac{m}{m-1} \bar{p} f^{(m-1)},  f^{(1)} = p.
\end{equation}
We split the sum using Pascal's binomial recurrence, massage the first term to $f^{(m-1)}$, and use the binomial absorbtion identity and binomial theorem for the second term:
\begin{eqnarray}
f^{(m)}
&=& \sum_{k=1}^m \frac{m}{k} \binom{m}{k} p^k \bar{p}^{m-k} \nonumber \\
&=& \sum_{k=1}^m \frac{m}{k} \left(\binom{m-1}{k}+\binom{m-1}{k-1}\right) p^k \bar{p}^{m-k} \nonumber \\
&=& \frac{m}{m-1} \bar{p}\sum_{k=1}^{m-1} \frac{m-1}{k} \binom{m-1}{k} p^k \bar{p}^{m-1-k} \nonumber \\
&+& \sum_{k=1}^m \frac{m}{k} \binom{m-1}{k-1} p^k \bar{p}^{m-k} \nonumber \\
&=& \frac{m}{m-1}\bar{p} f^{(m-1)} + \sum_{k=1}^m \binom{m}{k} p^k \bar{p}^{m-k}
\end{eqnarray}
Let $f = \lim_{m \to \infty} f^{(m)}$.  Taking limits on both sides of \eqref{exp1mrec} gives $f = 1 - \bar{p} f$, or $f=1/p$.  
\end{IEEEproof}

\begin{lemma}
\label{lem:varmMto0}
Let $M^{(m)} \sim \mathrm{bin}(m,p)$ be a binomial RV, and define $I^{(m)} = \frac{1}{M^{(m)}} \mathbf{1}_{M^{(m)}>0}$.  Then $m I^{(m)} \stackrel{p}{\to} 1/p$.
\end{lemma}

\begin{IEEEproof}
\lemref{expmMconv} gives $f^{(m)} = \Ebb[m I^{(m)}] \to \frac{1}{p}$.  It suffices to show $\mathrm{Var}(m I^{(m)}) \to 0$, or equivalently, since $\lim_{m \to \infty} (f^{(m)})^2 = \frac{1}{p^2}$ (by the continuous mapping theorem), to show $\lim_{m \to \infty} w^{(m)} = \frac{1}{p^2}$, where $w^{(m)} = \Ebb \left[(m I^{(m)})^2 \right]$.  We establish the recurrence
\begin{equation}
\label{eq:expmMsqrec}
w^{(m)} = \bar{p} \left( \frac{m}{m-1} \right)^2 w^{(m-1)} + f^{(m)}.
\end{equation}
As in the proof of the recurrence \eqref{exp1mrec} in the proof of \lemref{expmMconv}, we split the sum using Pascal's binomial recurrence, massage the first term into $w^{(m-1)}$, and use the binomial absorbtion identity and binomial theorem for the second term:
\begin{eqnarray}
w^{(m)}
&=& \sum_{k=1}^m \left(\frac{m}{k}\right)^2 \binom{m}{k} p^k \bar{p}^{m-k} \nonumber \\
&=& \sum_{k=1}^m \left(\frac{m}{k}\right)^2 \left(\binom{m-1}{k}+\binom{m-1}{k-1}\right) p^k \bar{p}^{m-k} \nonumber \\
&=& \bar{p} \left(\frac{m}{m-1}\right)^2 \sum_{k=1}^{m-1} \left(\frac{m-1}{k}\right)^2 \binom{m-1}{k} p^k \bar{p}^{m-1-k} \nonumber \\
&+& \sum_{k=1}^m \left(\frac{m}{k}\right)^2 \binom{m-1}{k-1} p^k \bar{p}^{m-k} \nonumber \\
&=& \bar{p} \left(\frac{m}{m-1}\right)^2 w^{(m-1)} + \sum_{k=1}^m \frac{m}{k} \binom{m}{k} p^k \bar{p}^{m-k}
\end{eqnarray}
Let $w = \lim_{m \to \infty} w^{(m)}$.  Taking limits on both sides of \eqref{expmMsqrec} gives $w = \bar{p}w + 1/p$, or $w=1/p^2$.  
\end{IEEEproof}

\begin{lemma}
\label{lem:covmMto0} 
Let $\Mbf^{(m)} = (M^{(m)}_1,\ldots,M^{(m)}_n) \sim \mathrm{mult}(m,\pbf)$ be a multinomial RV as in \eqref{multdisbn}, and let $a_i,a_j$ be constants.  Define $I^{(m)}_i = \frac{1}{M^{(m)}_i}\mathbf{1}_{M^{(m)}_i>0}$ and $I^{(m)}_j = \frac{1}{M^{(m)}_j}\mathbf{1}_{M^{(m)}_j>0}$.  Then $\mathrm{Cov}\left(a_i m I^{(m)}_i,a_j m I^{(m)}_j\right) \to 0$ as $m \to \infty$.
\end{lemma} 

\begin{IEEEproof}
By the definition of covariance and linearity of expectation, the constants $a_i,a_j$ are immaterial, and so without loss of generality we fix $a_i = a_j = 1$.  From \lemref{expmMconv} $\lim_{m \to \infty} \Ebb[m I^{(m)}] = \frac{1}{p}$.  Thus it suffices to show
\begin{equation}
\label{eq:lempfcov1mto0}
\lim_{m \to \infty} \Ebb \left[ m I^{(m)}_i  m I^{(m)}_j \right] = \frac{1}{p_i p_j}.
\end{equation}
For any pair of sequences of RVs, say $\{(V_1^{(m)},V_2^{(m)})\}$, with $V_1^{(m)} \stackrel{p}{\to} v_1$ and $V_2^{(m)} \stackrel{p}{\to} v_2$ as $m \to \infty$, it follows by Slutsky's Theorem that $(V_1^{(m)},V_2^{(m)}) \stackrel{p}{\to} (v_1,v_2)$ as $m \to \infty$, and then by the continuous mapping theorem that $V_1^{(m)}V_2^{(m)} \stackrel{p}{\to} v_1 v_2$ as $m \to \infty$.  Setting $V_1^{(m)} = m I^{(m)}_i$ and $V_2^{(m)} = m I^{(m)}_j$ gives \eqref{lempfcov1mto0}.
\end{IEEEproof}

We now proceed to the proof of \thmref{fairness}.
\begin{IEEEproof}
We employ the notation $I^{(m)}_i = \frac{1}{M^{(m)}_i}\mathbf{1}_{M^{(m)}_i>0}$, for $i \in [n]$.  The outline of the proof is as follows.  The random SE fairness $c(\Xbf^{(m)})$ is a ratio of RVs:
\begin{equation}
c(\Xbf^{(m)})= \frac{\left(\sum_{u=1}^m X^{(m)}_u \right)^2}{m\sum_{u=1}^m (X^{(m)}_u)^2} = \frac{N^{(m)}}{D^{(m)}}
\end{equation}
We will first prove
\begin{equation}
\label{eq:pffairfirststep}
N^{(m)} \stackrel{p}{\to} \left(\sum_{i=1}^n \psi_i \right)^2, ~
D^{(m)} \stackrel{p}{\to} \sum_{i=1}^n \frac{\psi^{(2)}_i}{p_i}, ~ m \to \infty.
\end{equation}
Given $N^{(m)} \stackrel{p}{\to} a$ and $D^{(m)} \stackrel{p}{\to} b$, Slutsky's Theorem ensures $(N^{(m)},D^{(m)}) \stackrel{p}{\to} (a,b)$, and the continuous mapping theorem guarantees $N^{(m)}/D^{(m)} \stackrel{p}{\to} a/b$, thereby proving the theorem.  It therefore remains to establish \eqref{pffairfirststep}.  We consider $N^{(m)}$ and $D^{(m)}$ in turn; first consider $N^{(m)}$.  From \corref{meanstddev} we have $\sum_{u=1}^m X^{(m)}_u \stackrel{p}{\to} \sum_{i=}^n \psi_i$.  Convergence in probability is preserved under continuous functions, and as such $N^{(m)} \stackrel{p}{\to} \left(\sum_{i=1}^n \psi_i \right)^2$, establishing the first part of \eqref{pffairfirststep}.  

Next consider $D^{(m)}$.  In what follows we drop the superscript $(m)$ to simplify the notation.  To establish $D \stackrel{p}{\to} \sum_{i=1}^n \frac{\psi^{(2)}_i}{p_i}$ it suffices to show $\Ebb[D] \to \sum_{i=1}^n \frac{\psi^{(2)}_i}{p_i}$ and $\mathrm{Var}(D) \to 0$. Consider in turn the mean (\underline{Step 1}) and variance (\underline{Step 2}) of $D$.  

\underline{Step 1:} $\Ebb[D]$.  Use $\Ebb[D] = \Ebb[\Ebb[D|\Mbf]]$, where 
\begin{eqnarray}
\Ebb[D|\Mbf] &=& m \sum_{i \in \Amc(\Mbf)} \sum_{u \in \Umc_i} \Ebb[X_u^2|M_i,i(u)=i] \nonumber \\
&=& m \sum_{i \in \Amc(\Mbf)} \sum_{u \in \Umc_i} \frac{\psi^{(2)}_i}{M_i^2} = \sum_{i=1}^n \psi^{(2)}_i m I_i \label{eq:pffairedgm1}
\end{eqnarray}
It follows that $\Ebb[D] = \sum_{i=1}^n \psi^{(2)}_i \Ebb \left[ m I_i \right]$ and \lemref{expmMconv} ensures $\Ebb[D] \to \sum_{i=1}^n \frac{\psi^{(2)}_i}{p_i}$ as $m \to \infty$.  

\underline{Step 2:} $\mathrm{Var}(D)$.  As in the proof of \thmref{meanstddev}, we use the notation $\Vrm(\cdot) = \mathrm{Var}(\cdot)$ and $\Crm(\cdot,\cdot) = \mathrm{Cov}(\cdot,\cdot)$, and the law of total variance:  
\begin{equation}
\label{eq:pffairltv1}
\Vrm(D) = \Ebb_{\Mbf}[\Vrm(D|\Mbf)] + \Vrm_{\Mbf}(\Ebb[D|\Mbf]).
\end{equation}
Consider the two terms in \eqref{pffairltv1} in turn: \underline{Step 2-1} and \underline{2-2}.  We will show both terms converge to $0$ as $m \to \infty$, and as such conclude $\Vrm(D) \to 0$.  

\underline{Step 2-1:} first term in \eqref{pffairltv1}: to find $\Ebb_{\Mbf}[\Vrm(D|\Mbf)]$ we first find 
\begin{equation}
\label{eq:pffairterm21}
\Vrm(D|\Mbf) = \Ebb[D^2|\Mbf] - \Ebb[D|\Mbf]^2 
\end{equation}
Consider the two terms in \eqref{pffairterm21} in turn: \underline{Step 2-1-1} and \underline{2-1-2}.  

\underline{Step 2-1-1:} first term in \eqref{pffairterm21}: $\Ebb[D^2|\Mbf]$ 
\begin{eqnarray}
&=& \Ebb\left[\left.\left(m \sum_{u=1}^m X_u^2 \right)^2 \right|\Mbf \right] \nonumber \\
&=& m^2 \Ebb\left[\left. \sum_{u=1}^m X_u^4 \right|\Mbf \right] + 2 m^2 \Ebb\left[\left. \sum_{1 \leq u < v \leq m} \!\!\!\!\! X_u^2 X_v^2 \right|\Mbf \right] \label{eq:pffair11a}
\end{eqnarray}
Consider the first term in \eqref{pffair11a}: $m^2 \Ebb\left[\left. \sum_{u=1}^m X_u^4 \right|\Mbf \right]$
\begin{eqnarray}
&=& m^2 \sum_{i \in \Amc(\Mbf)} \sum_{u \in \Umc_i} \Ebb[X_u^4 |M_i,i(u)=i] \nonumber \\
&=& \sum_{i=1}^n \psi^{(4)}_i \frac{m^2}{M_i^3} \mathbf{1}_{M_i > 0} = \frac{1}{m} \sum_{i=1}^n \psi^{(4)}_i (m I_i)^3.  
\end{eqnarray}
Consider the second term in \eqref{pffair11a} (without the $2 m^2$):
\begin{eqnarray}
&=& \sum_{i \in \Amc(\Mbf)} \sum_{\{u,v\} \in \Umc_i^2} \Ebb[X_u^2 X_v^2|M_i,i(u)=i,i(v)=i] \nonumber \\
&+& \!\!\!\!\! \sum_{\{i,j\} \in \Pmc(\Mbf)} \sum_{u \in \Umc_i} \sum_{v \in \Umc_j} \Ebb[X_u^2 X_v^2|M_i,M_j,i(u)=i,i(v)=j], \label{eq:pffairstep2110}
\end{eqnarray}
where $\Pmc(\Mbf)$ is the set of unordered pairs of occupied cells under $\Mbf$.  For any pair of users $\{u,v\}$ both in $\Umc_i$, $X_u^2,X_v^2$ are independent conditioned on $M_i$, and as such 
\begin{equation}
\label{eq:pffairstep211a}
\Ebb[X_u^2 X_v^2|M_i,i(u)=i,i(v)=i] = \left(\frac{\psi^{(2)}_i}{M_i^2} \right)^2.
\end{equation}
Similarly, for any pair of users $\{u,v\}$ with $u \in \Umc_i$ and $v \in \Umc_j$, $X_u^2, X_v^2$ are independent conditioned on $M_i,M_j$, and as such
\begin{equation}
\label{eq:pffairstep211b}
\Ebb[X_u^2 X_v^2|M_i,M_j,i(u)=i,i(v)=j] = \frac{\psi^{(2)}_i \psi^{(2)}_j}{M_i^2 M_j^2},
\end{equation}
Substitution of \eqref{pffairstep211a} and \eqref{pffairstep211b} into \eqref{pffairstep2110} gives that the second term in \eqref{pffair11a} (without the $2 m^2$) is 
\begin{equation}
\sum_{i=1}^n \binom{M_i}{2} \left(\frac{\psi^{(2)}_i}{M_i^2} \right)^2 \mathbf{1}_{M_i > 0} + \!\!\! \sum_{1 \leq i < j \leq n} \!\!\! M_i M_j \frac{\psi^{(2)}_i \psi^{(2)}_j}{M_i^2 M_j^2} \mathbf{1}_{M_i>0,M_j>0}.
\end{equation}
and thus we finish \underline{Step 2-1-1:} with 
\begin{eqnarray}
\Ebb[D^2|\Mbf] &=& \frac{1}{m} \sum_{i=1}^n \psi^{(4)}_i (m I_i)^3 + 2 \sum_{i=1}^n (\psi^{(2)}_i)^2 \binom{M_i}{2} m^2 I_i^4 \nonumber \\
&+& 2 \sum_{1 \leq i < j \leq n} \psi^{(2)}_i \psi^{(2)}_j m^2 I_i I_j \label{eq:pffairstep211end}
\end{eqnarray}

\underline{Step 2-1-2:} second term in \eqref{pffairterm21}: $\Ebb[D|\Mbf]^2$.  By \eqref{pffairedgm1}, $\Ebb[D|\Mbf]^2$
\begin{eqnarray}
&=& \left( \sum_{i=1}^n \psi^{(2)}_i m I_i \right)^2 \nonumber \\
&=& \sum_{i=1}^n (\psi^{(2)}_i)^2 m^2 I_i^2 + 2 \sum_{1 \leq i < j \leq n} \psi^{(2)}_i \psi^{(2)}_j m^2 I_i I_j \label{eq:pffairstep212end}
\end{eqnarray}

We now return to \underline{Step 2-1} by substituting \eqref{pffairstep211end} and \eqref{pffairstep212end} into \eqref{pffairterm21}, yielding: $\Vrm(D|\Mbf)$
\begin{equation}
=m^2 \sum_{i=1}^n \left( \psi^{(4)}_i I_i^3  - (\psi^{(2)}_i)^2 I_i^2 + 2 (\psi^{(2)}_i)^2 \binom{M_i}{2} I_i^4 \right) 
\label{eq:pffairstep21combine1}
\end{equation}
Simplifying gives
\begin{equation}
\Vrm(D|\Mbf) = \frac{1}{m} \sum_{i=1}^n \left( \psi^{(4)}_i - (\psi^{(2)}_i)^2 \right) \left(m I_i \right)^3.
\end{equation}
From \lemref{varmMto0} $m I_i \stackrel{p}{\to} \frac{1}{p_i}$, and thus by the continuous mapping theorem, $\left(m I_i\right)^3 \stackrel{p}{\to} \frac{1}{p_i^3}$.  It follows that $\Vrm(D|\Mbf) \stackrel{p}{\to} 0$, and as such $\Ebb_{\Mbf}[\Vrm(D|\Mbf)] \to 0$. 

\underline{Step 2-2:} second term in \eqref{pffairltv1}: to find $\Vrm_{\Mbf}(\Ebb[D|\Mbf])$ we use \eqref{pffairedgm1} and the equation for the variance of a sum of RVs: $\Vrm_{\Mbf}(\Ebb[D|\Mbf])=$
\begin{eqnarray}
\sum_{i=1}^n (\psi^{(2)}_i)^2 \Vrm \left( m I_i \right) + 2 \sum_{1 \leq i < j \leq n} \Crm\left(\psi^{(2)}_i m I_i,\psi^{(2)}_j m I_j \right)
\end{eqnarray}
By \lemref{varmMto0} the variances in the first sum converge to $0$, and by  \lemref{covmMto0} the covariances in the second sum converge to $0$.  It follows that $\Vrm_{\Mbf}(\Ebb[D|\Mbf]) \to 0$.
\end{IEEEproof}

\bibliographystyle{IEEEtran}
\bibliography{IEEEabrv,refs}
\end{document}